\documentclass[12pt,a4paper]{article}

\usepackage{amsmath}
\usepackage{amssymb}
\usepackage{epsfig}
\usepackage{graphics}

\def\nostrocostrutto#1\over#2{\mathrel{\mathop{\kern 0pt \rlap 
  {\raise.2ex\hbox{$#1$}}}
  \lower.9ex\hbox{\kern-.190em $#2$}}}

\catcode`@=11
\newcount\@tempcntc
\def\@citex[#1]#2{\if@filesw\immediate\write\@auxout{\string\citation{#2}}\fi
  \@tempcnta\z@\@tempcntb\m@ne\def\@citea{}\@cite{\@for\@citeb:=#2\do
    {\@ifundefined
       {b@\@citeb}{\@citeo\@tempcntb\m@ne\@citea\def\@citea{,}{\bf ?}\@warning
       {Citation `\@citeb' on page \thepage \space undefined}}%
    {\setbox\z@\hbox{\global\@tempcntc0\csname b@\@citeb\endcsname\relax}%
     \ifnum\@tempcntc=\z@ \@citeo\@tempcntb\m@ne
       \@citea\def\@citea{,}\hbox{\csname b@\@citeb\endcsname}%
     \else
      \advance\@tempcntb\@ne
      \ifnum\@tempcntb=\@tempcntc
      \else\advance\@tempcntb\m@ne\@citeo
      \@tempcnta\@tempcntc\@tempcntb\@tempcntc\fi\fi}}\@citeo}{#1}}
\def\@citeo{\ifnum\@tempcnta>\@tempcntb\else\@citea\def\@citea{,}%
  \ifnum\@tempcnta=\@tempcntb\the\@tempcnta\else
   {\advance\@tempcnta\@ne\ifnum\@tempcnta=\@tempcntb \else \def\@citea{--}\fi
    \advance\@tempcnta\m@ne\the\@tempcnta\@citea\the\@tempcntb}\fi\fi}
\catcode`@=12

\begin{document}




\setcounter{page}{0}
\thispagestyle{empty}
\begin{titlepage}

\vspace*{-1cm}
\hfill \parbox{3.5cm}{BUHE-00-09/ \\ BUTP-2000/26 \\ 
20. October, 2000 
\vspace*{0.3cm}
 }   
\vfill

\begin{center}
  {\large {\bf
Mapping out the QCD phase transition  \\
in multiparticle production} 
      \footnote{We thank the Schweizerische Nationalfonds for his support.}  }
\vfill
\vspace*{0.3cm} 

{\bf
    Sonja Kabana } \\
    Laboratory for High Energy Physics \\
    University of Bern \\
    CH - 3012 Bern, Switzerland
    \\
    E-mail: sonja.kabana@cern.ch
   \vspace*{0.3cm} \\  
   and \vspace*{0.3cm} \\
{\bf
    Peter Minkowski } \\
    Institute for Theoretical Physics \\
    University of Bern \\
    CH - 3012 Bern, Switzerland
    \\
    E-mail: mink@itp.unibe.ch
   \vspace*{0.3cm} \\  

\end{center}

\vfill

\begin{abstract}
\noindent
We analyze multiparticle production in a thermal framework for
7 central nucleus nucleus collisions, $e^+$+ $e^-$ annihilation into hadrons
on the Z resonance
and 4 hadronic reactions (p+p and p+$\overline{p}$ with partial centrality selection),
with center of mass energies ranging from $\sqrt{s}$= 2.6 GeV (per nucleon pair)
to 1.8 TeV. Thermodynamic parameters at chemical freeze-out (temperature and 
baryon and strangeness fugacities) are obtained from appropriate fits,
generally improving in quality for reactions subjected to centrality cuts.
All systems with nonvanishing fugacities are extrapolated along trajectories
of equal energy density, density and entropy density to zero fugacities.
The so obtained temperatures extrapolated to zero fugacities
as a function of initial energy density $\varepsilon_{\ in}$
universally show a strong rise followed by a saturating limit
of $T_{lim}$ = 155 $\pm$ 6 $\pm$ 20 MeV. We interpret this
behaviour as mapping out the boundary between quark gluon plasma and hadronic
phases. The ratio of strange antiquarks to light ones 
as a function of the initial energy density
$\varepsilon_{\ in}$ shows the same behaviour as the temperature,
saturating at a value of 0.365 $\pm$ 0.033 $\pm$ 0.07. 
No distinctive feature of
'strangeness enhancement' is seen for heavy ion collisions relative to 
hadronic and leptonic reactions, when compared at the same initial
energy density.
\end{abstract}

\vfill
\end{titlepage}


\newpage

\section{Introduction}

\noindent
Hadronic reactions involving copious production of secondary particles have
been associated with an underlying thermodynamic behaviour since the earliest observations in
cosmic rays \cite{rev}. The observable energy regime 
ranging up to 140 TeV per nucleon pair corresponding to incident cosmic ray
nucleons of $E \ \leq \ 10^{10} \ GeV$ demands elaborate simulations of the
observed extended air shower development , in order to extract definite multiplicity distributions
of the elementary hadronic and nuclear reactions \cite{capdevielle}.

\noindent
Thermodynamic models are widely and successfully used to describe identified particle ratios
in hadronic and especially heavy ion collisions \cite{Gerber} , \cite{revions},
\cite{revions1}, \cite{rafelski}, \cite{biro},
but were also extended to $e^+ e^-$ annihilation into hadronic final states 
\cite{becattee} , \cite{Chliapnikov}.
In heavy ion collisions two main parameters - energy per nucleon pair and centrality - 
are varied and their influence on thermodynamic variables - temperature
and chemical potentials - are studied.

\noindent
We derive and discuss thermal properties in search of the phase boundary 
between quark gluon plasma and condensed hadrons. We consider as one characteristic 
signature of
this boundary the critical dependence of kaon number densities on the
initial energy density.
The kaon multiplicities observed in Pb+Pb collisions at  
$\sqrt{s}$=17 GeV \cite{Na52}, 
and  in other nucleus+nucleus and p+p collisions at  
$\sqrt{s}$ $\sim$ 5-19 GeV \cite{hepph0004138,Sonja}, 
serve as motivation for our thermal description  extending
the previous analysis of one of us (SK) \cite{hepph0004138,Sonja}
(see figure \ref{t_k}). 

\noindent
In section 2 we discuss the dependence of the critical temperature
on the vaccum pressure in QCD, limited to the
case of vanishing chemical potentials, extending the work of reference
\cite{PM}.
In section 3 we seek to assign each thermodynamic state established at
chemical freeze-out for any given reaction for which chemical potentials
for baryon number and strangeness do not vanish,
an equivalent state at zero
fugacities. We do this extrapolating along curves of equal entropy density,
energy density and number density. 
The extrapolation of the temperature of systems with finite
fugacities to zero fugacities, and the investigation  of their
dependence on the initial energy density has been proposed in \cite{Sonja}.
We find that a more universal parameter as a measure of positive and negative strangeness 
production is $\lambda_{\ s} \ = \ \frac{2 \left \langle \overline{s} \right \rangle}
{ \left \langle \overline{u} \right \rangle \ + \ \left \langle \overline{d}  \right \rangle}$.
This parameter is also extrapolated to the zero fugacity systems. 

\noindent
The details of this procedure are worked out in the first four subsections
of section 3. The results at zero fugacities are contained in subsection
3.5 and represented in figures \ref{t} - \ref{t_k} in relation with the energy density
initially achieved in each reaction. 
Despite large errors, the phase boundary between
quark gluon plasma and hadronic phases is clearly mapped out.

\noindent
We use in the hadronic phase the approximation of noninteracting hadron resonances
to describe in this sense global
ratios of hadrons produced in the following reactions :

\begin{description}

\item 1) \begin{tabular}[t]{l}
central Au+Au collisions at RHIC $\sqrt{s} \ = 130 \ A \ GeV$
\cite{hepex0007036} , \cite{star} .
\end{tabular}

\item 2) \begin{tabular}[t]{l}
central Pb+Pb collisions at SPS at $\sqrt{s} \ = 17 \ A \ GeV$ 
\cite{nuclth9903010} , \\
central S+A collisions ( A=Au,W,Pb) at SPS  
$\sqrt{s} \ = 19 \ A \ GeV$ \cite{nuclth9508020} , \\

central S+S collisions at SPS $\sqrt{s} \ = 19 \ A \ GeV$ 
\cite{na35_papers} .
\end{tabular}

\item 3) \begin{tabular}[t]{l}
central Si+Au collisions at AGS 
$\sqrt{s} \ = 5.4 \ A \ GeV$ \cite{revions1} ,\\

central Au+Au collisions at AGS  
$\sqrt{s} \ = 4.9 \ A \ GeV$ \cite{Braun-M}, \cite{Cley}.
\end{tabular}

\item 4) \begin{tabular}[t]{l}
Ni+Ni collisions at GSI $\sqrt{s} \ = 2.8 \ A \ GeV$ \cite{fopi,senger_stroebele} .
\end{tabular}

\item 5) \begin{tabular}[t]{l}
$e^+$+ $e^-$ collisions at LEP  $\sqrt{s} \ = 91.19 \ GeV$
\cite{becattee}, \cite{Chliapnikov} .
\end{tabular}

\item 6) \begin{tabular}[t]{l}
p+p collisions at $\sqrt{s} \ = \ 27 \ GeV$ \cite{hepph9702274} , \\
p+p collisions at $\sqrt{s} \ = \ 17 \ GeV$ \cite{siklerqm99} , \cite{na49phi} .
\end{tabular}

\item 7) \begin{tabular}[t]{l}
$p+\overline{p}$ collisions at $\sqrt{s} \ = \ 900 \ GeV$ \cite{hepph9702274} , \\
peripheral $p+\overline{p}$ collisions at $\sqrt{s} \ = \ 1.8 \ TeV$
\cite{Gutay} , \\
central $p+\overline{p}$ collisions at $\sqrt{s} \ = \ 1.8 \ TeV$ \cite{Gutay} .
\end{tabular}

\end{description}
\newpage

\section{Outline of basic points}

\noindent
We consider first the thermodynamic potential $\Phi$ of a grand canonical ensemble of
hadron resonances without further interactions among them. 
The thermodynamic variables are  

\begin{equation}
\label{eq:1}
\begin{array}{l}
\begin{array}{llll}
V, & T, & \chi_{\ B} \ = \ \mu_{\ b} \ / \ T, & \chi_{\ s} \ = \ \mu_{\ S} \ \ / T 
\end{array}
\end{array}
\end{equation}

\noindent
i.e. volume, temperature, baryon and strangeness fugacity.
The fugacity of the third component of isospin is set to zero,
neglecting isospin asymmetries.
\vspace*{-0.3cm}

\begin{equation}
\label{eq:2}
\begin{array}{l}
\Phi \ = \ g \ V
\hspace*{0.1cm} ; \hspace*{0.1cm}
g \ = \ g \ ( \ \beta \ , \ \chi_{\ b} \ , \ \chi_{\ s} \ ) \ =
\ \sum_{\ \alpha} \ g_{\ \alpha} \ + \ g_{\ 0}    
\vspace*{0.3cm} \\
\beta = \ 1 \ / \ T
\hspace*{0.1cm} ; \hspace*{0.1cm}
g \ = \ \beta \ p
\hspace*{0.2cm} ; \hspace*{0.2cm}
p \ : \ \mbox{pressure}
\end{array}
\end{equation}

\noindent
In eq. (\ref{eq:2}) the sum extends over hadron resonances denoted by $\alpha$,
where we only include the pseudoscalar and vector u,d,s meson nonets as well
as the spin 1/2 baryon octet and spin 3/2 decuplet and their
antiparticles as well as the $f_0(400-1200)$ or $\sigma$, interpreted
as scalar glueball \cite{SKPM} for simplicity.

\noindent
$g_{\ 0} \ = \ \beta \ p_{\ 0}$ takes into account the nonzero vacuum pressure
characterizing the hadronic phase in QCD, which we restrict to the 
three light flavors. 
\vspace*{-0.3cm}

\begin{equation}
\label{eq:3}
\begin{array}{l}
p_{\ 0} \ ( \ T \ = \ 0 \ ) \ = 
\ \begin{array}{c}
 9
 \vspace*{0.3cm} \\
\hline	\vspace*{-0.3cm} \\
32 \ \pi^{\ 2}
\end{array}
\ {\cal{B}}^{\ 2} \ - \ \frac{1}{4} \ \Lambda \ ( \ m_{\ q} \ )
\ =
\ \left \lbrace
\begin{array}{l}
0.00302 \ GeV^{\ 4}
 \vspace*{0.3cm} \\
0.00658 \ GeV^{\ 4}
\end{array} \right .
 \vspace*{0.3cm} \\
{\cal{B}}^{\ 2} \ =
\ \left \langle \ 0 \ \right |
\ \frac{1}{4} \ F_{\ \mu \nu}^{\ a} \ F^{\ a \ \mu \nu}
\ \left | \ 0 \ \right \rangle 
\ =
\ \left \lbrace
\begin{array}{l}
0.125 \ GeV^{\ 4} \ \cite{SVZ}
 \vspace*{0.3cm} \\
0.250 \ GeV^{\ 4} \ \cite{Nari}
\end{array} \right .
 \vspace*{0.3cm} \\
\Lambda  =  \sum_{\ q} \ m_{\ q} 
\ \left \langle \ 0 \ \right |
 - \ \overline{q} \ q
\ \left | \ 0 \ \right \rangle 
 \sim \ f_{\ \pi}^{\ 2} \ ( \ \frac{1}{2} \ m_{\ \pi}^{\ 2} \ + \ m_{\ K}^{\ 2} \ )
\ = 0.00217 \ GeV^{\ 4}
\end{array}
\end{equation}

\noindent
Converting energy density units to $GeV \ / \ fm^{\ 3}$ the range of values
for the (positive) vacuum pressure becomes

\begin{equation}
\label{eq:4}
\begin{array}{l}
p_{\ 0} \ ( \ T \ = \ 0 \ ) \ = 
\ \left \lbrace
\begin{array}{l}
0.377 \ GeV \ / \ fm^{\ 3}
 \vspace*{0.1cm} \\
0.823 \ GeV \ / \ fm^{\ 3}
\end{array} \right .
 \vspace*{0.3cm} \\
p_{\ 0} \ ( T ) \ \sim 
\ p_{\ 0} \ ( T=0 )  
\ \left ( \ 1 \ - \ ( T / T_{\ cr} )^{\ 4} \ \right )
\end{array}
\end{equation}

\noindent
The temperature variation of the vacuum pressure is an approximation
with $T_{\ cr} \ = \ T_{\ cr} \ ( \ \chi_{\ b} \ , \ \chi_{\ s} \ )$.

\noindent
The quantities $g_{\ \alpha}$ in eq. (\ref{eq:2}) for noninteracting resonances
are then given by 

\begin{equation}
\label{eq:5}
\begin{array}{l}
g_{\ \alpha} \ ( \ \beta \ , \ \chi_{\ b} \ , \ \chi_{\ s} \ ) \ =
\ w_{\ \alpha} \ {\displaystyle{\int}}_{\ m_{\ \alpha}}^{\ \infty} 
\ dE \ E \ p \ / \ ( \ 2 \pi^{\ 2} \ ) \ l  
 \vspace*{0.3cm} \\
 l \ = \ \mp \ \log \ \left \lbrack \ 1 \ \mp
 \ \exp \ ( \ - \ \beta \ E \ + \ \chi_{\ b} \ B_{\ \alpha}
 \ + \ \chi_{\ s} \ S_{\ \alpha} \ ) \ \right \rbrack
 \vspace*{0.3cm} \\
w_{\ \alpha} \ = \ ( \ 2 \ I_{\ \alpha} \ + \ 1 \ )
\ ( \ 2 \ Sp_{\ \alpha} \ + \ 1 \ )
\hspace*{0.2cm} ; \hspace*{0.2cm}
p \ = \ \sqrt{ E^{\ 2} \ - \ m_{\ \alpha}^{\ 2}}
\end{array}
\end{equation}

\noindent
In eq. (\ref{eq:5}) $( \ I \ , \ Sp \ , \ B \ , \ S \ )_{\ \alpha}$
denote isospin, spin, baryon number and strangeness of the hadron resonance $\alpha$.
The -,+ signs apply to bosons and fermions respectively.

\noindent
If we take into account the variation of the vacuum pressure with 
temperature then the masses 
$m_{\ \alpha} \ \rightarrow \ m_{\ \alpha} \ ( \ T \ )$ of quasiexcitations become
temperature dependent quantities. 

\noindent
In all subsequent calculations we neglect these temperature dependent 
effects, which however do set in dramatically, when the temperature
deviates from the crirtical one by less than 10 \%.

\begin{equation}
\label{eq:6}
\begin{array}{l} 
p_{\ 0} \ ( \ T \ ) \ \rightarrow \ p_{\ 0} \ =
\ p_{\ 0} \ ( \ T=0 \ )
\hspace*{0.2cm} ; \hspace*{0.2cm}
m_{\ \alpha} \ ( \ T \ ) \ \rightarrow \ m_{\ \alpha} \ =
m_{\ \alpha} \ ( \ T=0 \ )
\end{array}
\end{equation}

\noindent
We state here, that a thermal description cannot account
for any azimuthal ~( $\varphi$ - ) dependence of inclusive cross sections.
Thus azimuthal anisotropies  provide a measure for  non-equilibration.
\vspace*{0.1cm}

\noindent
The potential $\Phi$ in eq. (\ref{eq:2}) and the entropy denoted by
${\cal{S}}$ give rise to
the differentials

\begin{equation}
\label{eq:7}
\begin{array}{l} 
d \ \Phi \ =
\ - \ {\cal{E}} \ d \ \beta \ + \ N_{\ \nu} \ d \ \chi_{\ \nu}
\ + \ g \ d \ V
 \vspace*{0.3cm} \\
d \ {\cal{S}} \ = 
\ + \ \beta \ d \ {\cal{E}} \ - \ \chi_{\ \nu} \ d \ N_{\ \nu}
\ + \ g \ d \ V
\hspace*{0.2cm} \rightarrow \hspace*{0.2cm}
 \vspace*{0.3cm} \\
{\cal{S}} \ = \ \varrho_{\ s} \ V \ = \ \Phi \ + \ \beta \ {\cal{E}} 
\ - \ \chi_{\ \nu} \ N_{\ \nu}
\hspace*{0.2cm} ; \hspace*{0.2cm}
\nu \ = \ b \ , \ s
 \vspace*{0.3cm} \\
\varrho_{\ s} \ = \ \beta \ ( \ \varepsilon \ + \ p \ ) 
\ - \ \chi_{\ \nu} \ \varrho_{\ \nu}
\end{array}
\end{equation}

\noindent
In eq. (\ref{eq:7}) $\varrho_{\ s}$ denotes the entropy density, $\varepsilon$
the energy density and $\varrho_{\ b,s}$ the net baryon and strangeness
(number) density respectively.

\begin{equation}
\label{eq:8}
\begin{array}{l} 
\varrho_{\ \nu} \ = \ \sum_{\ \alpha} \ \nu_{\ \alpha} \ \varrho_{\ \alpha}
\hspace*{0.2cm} ; \hspace*{0.2cm}
\nu_{\ \alpha} \ = \ \left ( \ B_{\ \alpha} \ , \ S_{\ \alpha} \ \right )
\end{array}
\end{equation}

\noindent
As is apparent from eq. (\ref{eq:7}) the entropy density obtains no contribution
from vacuum energy density and pressure, as long as the vacuum
pressure is taken as independent of temparature as indicated in
eq. (\ref{eq:6}).

\noindent
Energy density and total number density then take the form

\begin{equation}
\label{eq:9}
\begin{array}{l} 
\varepsilon \ = \ \varrho_{\ e} \ - \ p_{\ 0}
\hspace*{0.2cm} ; \hspace*{0.2cm}
\varrho_{\ e} \ = \ \sum_{\ \alpha} \ \varrho_{\ e \alpha}
\hspace*{0.2cm} ; \hspace*{0.2cm}
\varrho_{\ n} \ = \ \sum_{\ \alpha} \ \varrho_{\ n \alpha}
 \vspace*{0.3cm} \\
\left \lbrace
\ \begin{array}{l}
\varrho_{\ e \alpha}
 \vspace*{0.2cm} \\
\varrho_{\ n \alpha}
\end{array}
\ \right \rbrace
\ = 
\ w_{\ \alpha} \ {\displaystyle{\int}}_{\ m_{\ \alpha}}^{\ \infty} 
\ dE \ E \ p \ / \ ( \ 2 \pi^{\ 2} \ )  
\ \left \lbrace
\ \begin{array}{l}
E
 \vspace*{0.2cm} \\
1
\end{array}
\ \right \rbrace
\ n
 \vspace*{0.3cm} \\
n \ =
\ \begin{array}{c}
 1
 \vspace*{0.3cm} \\
\hline	\vspace*{-0.3cm} \\
  \exp \ ( \ \beta \ E \ - \ \chi_{\ b} \ B_{\ \alpha}
 \ - \ \chi_{\ s} \ S_{\ \alpha} \ ) \ \mp \ 1
\end{array}
\end{array}
\end{equation}

\noindent
The approximation of hadronic interactions by free hadron resonances
is of course at best a consistent approximate thermodynamic description.
We will use it here to describe in this sense global
ratios of hadrons produced in the reactions listed in the introduction.

\noindent
We will present a detailed analysis of these collisions, as seen
at the time of chemical freeze-out in the next section.
The key parameter characterizing not this freeze-out instance but
the time of first onset of thermal conditions is the initial
energy density $\varepsilon_{in}$. It determines
the thermal parameters of the hadronic system as seen at chemical
freeze-out and distinguishes those systems which have hadronized
after transcurring the quark gluon plasma phase
from those which remained throughout in the hadronic phase.
\vspace*{0.1cm}

\noindent
{\bf The phase boundary for vanishing fugacities}
\vspace*{0.1cm}

\noindent
In the remainder of this section we analyze the equilibrium condition
for both phases to coexist in the case of vanishing fugacities 
or equivalently chemical potentials. This latter choice is just
for clarity of argument and also because for limiting center of 
mass energies and finite baryon number, strangeness and charge of
the initial system this is the limiting case. 
 
\noindent
This condition corresponds to equating the Gibbs densities of the
two phases

\begin{equation}
\label{eq:10}
\begin{array}{l} 
g^{\ H} \ ( \ T \ ) \ =
\ g^{\ QGP} \ ( \ T \ ) 
\hspace*{0.2cm} ; \hspace*{0.2cm}
\chi_{\ \nu} \ = \ 0
\hspace*{0.2cm} \rightarrow \hspace*{0.2cm}
T \ = \ T_{\ cr}
\vspace*{0.3cm} \\ 
g^{\ H} \ = \ \sum_{\ \alpha} \ g_{\ \alpha} \ + \ \beta \ p_{\ 0}    
\hspace*{0.2cm} ; \hspace*{0.2cm}
g^{\ QGP} \ = \ \sum_{\ - \alpha} \ g_{\ - \alpha} 
\vspace*{0.3cm} \\ 
\pm \ \left \lbrace \ \alpha \ \right \rbrace
\ =
\ \left \lbrace
\begin{array}{l} 
+ \ : \ \mbox{hadron resonances}
\vspace*{0.2cm} \\ 
- \ : \ \mbox{gluons, quarks and antiquarks (u,d,s)}
\end{array}
\ \right .
\end{array}
\end{equation}

\noindent
In the same approximative vein as applied to the hadronic phase (H)
we approximate the various quasiexcitations in the quark gluon
phase by free gluons, quarks and antiquarks analogous to
the form of $g_{\ \alpha}$ in eq. (\ref{eq:5})

\begin{equation}
\label{eq:11}
\begin{array}{l} 
g_{\ - \alpha} \ = 
\ w_{\ - \alpha} \ {\displaystyle{\int}}_{\ m_{\ - \alpha}}^{\ \infty} 
\ dE \ E \ p \ / \ ( \ 2 \pi^{\ 2} \ ) \ l  
\vspace*{0.3cm} \\ 
 l \ = \ \mp \ \log \ \left \lbrack \ 1 \ \mp
 \ \exp \ ( \ - \ \beta \ E \ ) \ \right \rbrack
\hspace*{0.2cm} ; \hspace*{0.2cm}
p \ = \ \sqrt{ E^{\ 2} \ - \ m_{\ - \alpha}^{\ 2}}
\vspace*{0.3cm} \\ 
- \alpha
\hspace*{0.2cm} : \hspace*{0.2cm}
\ \left \lbrace
\begin{array}{lll l} 
gl \ : & w_{\ gl} \ = \ 16 & m_{\ gl} \ = \ 0 & -
\vspace*{0.2cm} \\ 
q \ : & w_{\ q} \ = \ 6 & m_{\ q} & +
\vspace*{0.2cm} \\ 
\overline{q} \ : & w_{\ \overline{q}} \ = \ 6 & m_{\ q} & +
\end{array}
\right .
\vspace*{0.3cm} \\ 
m_{\ u, d, s} \ = 
\ \left ( \ 5.25 \ , \ 8.75 \ , \ 175 \ \right ) \ MeV
\end{array}
\end{equation}

\noindent
In eq. (\ref{eq:11}) the color/spin weights, masses and boson/fermion
signs are specified. The quark masses represent
a renormalization group invariant 'best' choice. 
It turns out that near the critical temperature 
$T_{\ cr} \ = \ O \ ( \ 200 \ MeV \ )$ the (anti)quark contributions
to the Gibbs density do not deviate substantially from the
limit of vanishing quark masses including $m_{\ s}$.

\begin{figure}[ht]
\begin{center}
\mbox{\epsfig{file=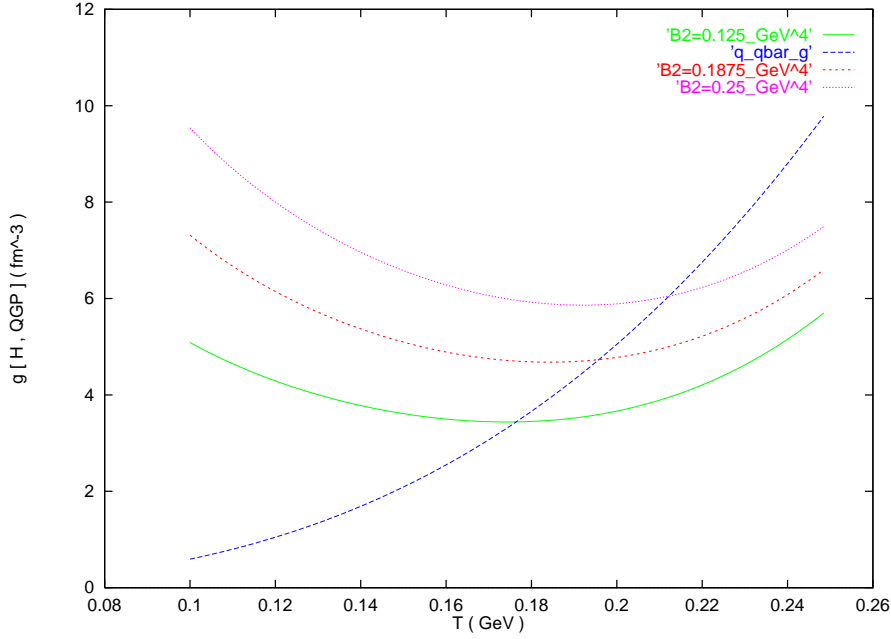,width=120mm}}
\end{center}
\caption{
     Gibbs potentials as a function of the temperature for three values
     of the gluon condensate in the ground state, 
     and for free quark flavors u, d, s and gluons.
     }
     \label{gbfig3}
     \end{figure}

\noindent
The crossings of $g^{\ H}$ and $g^{\ QGP}$ as displayed in figure \ref{gbfig3}
seem to suggest a first order phase transition with respect to energy density.

\noindent
This is however due to the approximation of fixed masses for hadron excitations
and of free quark and gluon flavors in the region of $T_{\ cr} \ = \ 194 \ \pm \ 18$ MeV.
We note that the above estimate of the critical tempeature at zero fugacities
confirms within the theoretical error the previous estimate of one of us \cite{PM}.

\noindent
Within the approximations and errors we obtain the critical temperature
and critical hadronic energy density on the side of the hadronic phase
for vanishing fugacities

\begin{equation}
\label{eq:12}
\begin{array}{l} 
T_{\ cr} \ = \ 194 \ \pm \ 18 \ MeV
\hspace*{0.2cm} ; \hspace*{0.2cm} 
\varrho_{\ e \ cr} 
 \left (
 \begin{array}{l}
212 \ MeV \\
194 \ MeV \\
176 \ MeV
\end{array}
 \right )
\ = 
\ \left \lbrace
\begin{array}{l} 
1.797 \\
1.018 \\
0.5406
\end{array}
\ \right .
 GeV/fm^{3}
\end{array}
\end{equation}

\noindent
Obviously the critical boundary curve could be extended to arbitrary chemical
potentials, but again the 'melting' of masses of hadronic quasiexcitations
modifies the Gibbs density on the side of the hadronic phase, whereas
the approximation of free quark and gluon modes does not warrant the
extrapolation to arbitrary chemical potentials, especially for small
temperature.
\newpage

\section{Reduction of chemical freeze-out parameters to zero fugacities 
in multiparticle production}

\noindent
In this section we apply the thermal model introduced in the
last section (equation 5) to extract the intensive thermodynamic parameters
defined in eq. 1.
We then extrapolate these parameters to zero fugacities along
states with equal entropy density, energy density or density.
The goal of this analysis is to compare the temperature at zero fugacities
with the initial energy density achieved in the collision in order to reveal
a boundary reflecting the QCD phase transition.
\vspace{0.3cm}

\noindent
One serious well known problem when comparing models to experimental data is
due to decays of resonances. This effect is called 'feeding'.
We compare calculated particle ratios to experimental data,
taking all (strong, electromagnetic and weak) 
decays with a branching ratio not below 1\% into account.
We try to account for experimental acceptance for $K^+$ and $K^-$, assuming
a 50\% feeding to pions, due to their long decay length.
The other weak decays (e.g. $\Lambda$, $K^0_s$) have a much shorter
decay length and are assumed to fully feed into secondary particles.
The $K^0_l$ decay is not considered.
\vspace{0.3cm}

\noindent
Contrary to references \cite{revions,revions1} we do not take all hadronic
resonances below 2 GeV mass into account.
The theoretical error inherent to the free resonance approximation allows
in our opinion our reduced set.
This is supported by the finding that we derive similar thermodyamic parameters
with similar accuracy as e.g. the ones found in reference \cite{nuclth9903010}.
Though the high mass resonances are definitely produced at sufficiently
high energy, it is not clear that the thermal description becomes more accurate
by including them because the quasiexcitations may not correspond to them,
especially if these are quark and gluon modes.
All similar thermal descriptions suffer from replacing interactions by noninteracting
resonances.
Furthermore, the isotropic angular distributions characterizing the models
are far from reality \cite{qmconference,flow} (e.g. flow phenomena, diffractive phenomena).

\vspace{0.3cm}
\noindent
We enforce strangeness conservation, to obtain the best fit,
exept in the case of p+p collisions at $\sqrt{s}$=17 GeV.
We find the error on the fitted parameters, at $\pm$ 1 unit of
the ratio of $\chi^2/DOF$ away from the best fitted value.
Our quoted error reflects the inherent theoretical error,
despite the difficulty to quantify it.
Note that other analyses \cite{revions,revions1} finding a 
smaller error seem to use as error the fitted parameters at $\pm$ 1 unit of
the $\chi^2$ itself.

\vspace{0.3cm}
\noindent
In the following sections we note if the data we use
are taken with a minimum bias
trigger (that means no trigger bias is imposed) or a central respectively a 
peripheral trigger.
Central is here understood as the lowest impact parameter region.
It is usually selected in the experiment by considering only collisions 
 with the largest particle
 multiplicity (e.g. CERN-WA97, Fermilab-E735)
 or the largest transverse energy (e.g. CERN-NA52),
or the smallest forward going energy (which reflects
the spectator nucleon number) (e.g. CERN-NA49).
Respectively, peripheral collisions are a selection of collisions
with the largest impact parameters using the same means as described
above (charged multiplicity, energy).

\vspace{0.3cm}

\noindent
In the following analysis we also give the quantity:
$
\lambda_s \ = \ 
\frac{ (2 \overline{s}) }
{ (\overline{u} + \overline{d}) }
$
which we take as a measure of the strangeness suppression factor 
defined and used in the literature as
$
\lambda_s \ = \ 
\frac{ 2 (s + \overline{s}) }
{ (u + \overline{u} + d + \overline{d}) }
$
\cite{revions1}.
We use antiquarks to consider in a simplified way only the newly produced valence antiquarks.
We give also the equivalent $ \lambda_s$ at zero fugacities.
From quark counting rules one expects $ \lambda_s$  to be in the range 0.3-0.5
\cite{quark_rules}.
In principle one should include in any definition of $ \lambda_s$ the newly 
produced sea quarks however the relevant proportion of the latter inside
hadrons, is still subject of
experimental and theoretical investigations.
\vspace{0.3cm}

\noindent
Generally we find that thermal fits to minimum bias or peripheral nuclear and 
hadronic reactions are not satisfactory, in contrast to central
collisions.
We conjecture that this is due to the presence of at least two thermal sources
which we attribute to diffractive versus pomeron induced subprocesses.
Diffractive processes at finite $\sqrt{s}$
feed back into the midrapidity region, whereas pomeron induced  subprocesses 
result in particle production at midrapidity dominantly.
Following this conjecture, we conclude that the relative importance of 
diffractive to pomeron induced  subprocesses
decreases with increasing  centrality of the collision.
In this respect it would be helpful firstly, to increase experimental coverage
in the target and/or the projectile fragmentation regions.
This study has been proposed by \cite{bjorken,pschlein}.
Secondly, it is important that experiments 
apply centrality cuts, in nucleus+nucleus as well as in
p+nucleus and in elementary particle collisions.

\vspace{0.5cm}
\vspace{0.5cm}
\vspace{0.5cm}
\vspace{0.5cm}

\subsection{Nucleus nucleus reactions}

\noindent
In the following 4 subsections we analyse data from 
nucleus nucleus collisions at 1. RHIC, 2. SPS, 3. AGS and 4. SIS energies.

\vspace{0.5cm}
\vspace{0.5cm}

\subsubsection{Central Au+Au collisions at $\sqrt{s}$=130 GeV}

\noindent
We used 3 measured ratios at midrapidity 
from  \cite{hepex0007036,star} :
$\overline{ \Lambda} / \Lambda $, $\overline{p}/p$ and $h^-$/ Ncharged
  and imposed strangeness conservation. 
  We exclude weak decay products as is done in the experiment 
  \cite{hepex0007036}.
  The data from reference \cite{star} are preliminary.
The predicted 
and the experimental ratios are shown in table \ref{ratios_auau130gev}.
The resulting $\chi^2/DOF$ is (1.41/1) (CL 23\%) for a temperature of
168 $\pm$ 40 MeV.
When adding a systematic error of 15\% linearly to account for the
experimental feeding uncertainties and the fact that the ratios are measured at
midrapidity only,
 the resulting $\chi^2/DOF$ is 0.27/1 (CL 60\%).
The errors on the temperature and $\lambda_s$
are not well estimated in this analysis due to the
insensitivity of the ratios used, to the temperature at fixed fugacities.
We estimate the error on the temperature searching for
a variation of the $\chi^2/DOF$ by 1, while changing the fugacities
and fixing each time the T by imposing strangeness conservation.
In order to improve the fit quality the addition of other ratios
more sensitive to temperature variations  is needed,
like $K/\pi$ etc., which will soon be measured by the  experiments at RHIC. 
 \\
After defining the (T, $\mu_b$, $\mu_s$) values describing the particle
ratios produced in central Au+Au collisions at 
$\sqrt{s}$=130 GeV at the chemical freeze-out 
we extrapolate to the T at zero fugacities (table \ref{t_auau130gev}).
Because of the very small value of the fugacities, the extrapolated
values do not differ much from the ones at finite fugacitites.
\\
The $\chi^2/DOF$ of all reactions are shown in table \ref{chisquare}.

\begin{table}[ht]
\begin{tabular}{|lll|}
\hline
ratio  & model & data  \\
\hline

$\overline{ \Lambda} / \Lambda$  & 0.718 & 0.70 $\pm$ 0.25 \\

$\overline{p}/p$  & 0.662 & 0.650 $\pm$ 0.090\\

$h^-/Nch$ & 0.491  & 0.432 $\pm$ 0.0504 \\

\hline
\end{tabular}
\caption{
RHIC Au+Au at $\sqrt{s}$= 130 GeV. \protect\newline
Predicted versus experimental particle ratios for the best fit of our model.
}
\label{ratios_auau130gev}
\end{table}

\begin{table}[ht]
\begin{tabular}{|lllllll|}
\hline
$\mu_b$   &  $\mu_s$     &  T   & $\lambda_s$ &  
$\rho_e$  &  $T_{eq, \rho_e}$   & $\lambda_s({eq, \rho_e})$   \\
GeV       &  GeV         &  GeV &             &
GeV/fm$^3$ & GeV               &     \\
\hline

0.0390     &   0.0084   &  0.168  & 0.443 
& 0.401    &   0.169   &  0.422  \\

    &     &   $\pm$ 0.040 &  
&    &    $\pm$ 0.040 &  +0.105 -0.156 \\

\hline
  &   &    &   &
  $\rho_n$  & $T_{eq, \rho_n}$  &  $\lambda_s({eq, \rho_n})$   \\
       &           &   &             &
1/fm$^3$ & GeV               &     \\
\hline


  &    &    & 
&  0.453  &   0.168   & 0.443  \\

  &    &    & 
&    &    $\pm$ 0.040  &  +0.11 - 0.164 \\

\hline
  &   &    &   &
$\rho_s$  & $T_{eq, \rho_s}$  &   $\lambda_s({eq, \rho_s})$    \\
       &           &   &             &
1/fm$^3$ & GeV               &     \\
\hline


  &    &    & 
&  2.83  & 0.168  &  0.443  \\

  &    &    & 
&   &  $\pm$ 0.040  &   +0.11 - 0.164  \\

\hline
\end{tabular}
\caption{ 
RHIC Au+Au at $\sqrt{s}$= 130 GeV. \protect\newline
Thermodynamic parameters for the best fit and temperatures
and $\lambda_{\ s}$
extra\-polated to zero fugacities.
%
}
\label{t_auau130gev}
\end{table}

\clearpage

\subsubsection{Central nucleus nucleus collisions at $\sim$ 200 A GeV}

\noindent
Here we show results from central  Pb+Pb collisions at 158 A GeV,
and from S+S and S+A collisions at 200 A GeV.
\vspace{0.5cm}

\noindent{ \bf Central Pb+Pb collisions at 158 A GeV} \\
\vspace{0.5cm}

\noindent
For central  Pb+Pb collisions at 158 A GeV,
we use the particle ratios from table I in \cite{nuclth9903010} to compare with
our model predictions.
We don't consider a chemical potential for isospin \cite{nuclth9903010}, 
therefore we don't
consider e.g. the $\pi^- / \pi^+$ ratio.
We did not use the ratios in \cite{nuclth9903010}  for which feeding correction is included.
We did not use the $\overline{d}/d$ ratio, since there is strong experimental
evidence that $d$ and $\overline{d}$ are formed in the thermal hadronic freeze-out through
coalescence \cite{Na52,heinzscheibl}.
Furthermore, the $\overline{p}$ may not freeze out in the chemical hadronic freeze-out
due to the high cross section for $p + \overline{p}$ annihilation to e.g. $\pi$ as
noted in \cite{shuryak_ppbar}.

\noindent
We introduce a systematic error of 14\% (quadratically added) in the ratios including WA97 data, 
because they are measured at midrapidity only. 
We add also quadratically) a 10\% systematic error to ratios including $h^-$
due to uncertainty in the feeding correction.
The predicted ratios are
shown in table \ref{ratios_pbpb} together with the experimental data.
\vspace*{0.3cm}

\noindent
The $\chi^2$ at which strangeness is conserved is 16.1 over 12 degrees of freedom 
(CL=20 \% $\chi^2/DOF$=1.34) and corresponds to T=162 +10 -28 MeV.
The asymmetry of the error (defined as the T deviation 1 unit of $\chi^2/DOF$
away)
is due to the fact that the 16.1 is not at the minimum.
The minimum of the $\chi^2$ is 11.9 over 11 DOF (CL=50\%, $\chi^2/DOF$=1.08) 
at T=154 +14 -16 MeV.
There is a probable physics process
which may induce an imbalance of strangeness in experiment,
if the measurements do not include the target and beam fragmentation regions:
in particular particles with negative $S$ especially hyperons may be produced 
near beam rapidity.
In ref. \cite{rafelski_ssbar_imbalance} it is conjectured that strangeness is not
balanced (with missing particles with $S$ negative).
However we will demand in the following exact strangeness conservation.

\noindent
After defining the (T, $\mu_b$, $\mu_s$) values describing the particle
ratios produced in central Pb+Pb collisions at 158 A GeV at the chemical freeze-out 
we extra\-polate to T at zero fugacities.
The resulting equivalent temperatures for equal entropy density ($\rho_s$),
energy density ($\rho_e$), and number density ($\rho_n$) 
are shown in table \ref{t_pbpb}.

\begin{table}[ht]
\begin{tabular}{|l|l|l|}
\hline
ratio  & model & data  \\
\hline

$(p - \overline{p}) / h^- $    
&  0.165  &  0.228 $\pm$ 4.31E-2 \\





$\eta/ \pi^0$   &  7.35E-2 & 8.10E-2 $\pm$  1.30E-2 \\

$K^0_s / \pi^-$   &  0.142 & 0.125 $\pm$ 1.90E-2 \\

$K^0_s / h^-$	 &  0.124  &  0.123 $\pm$ 2.64E-2 \\

$ \Lambda/ h^-$  &  1.04E-1 & 7.70E-2 $\pm$ 1.54E-2 \\

$ \Lambda/ K^0_s $   & 0.835 & 0.630 $\pm$ 1.02E-1 \\

$K^+ / K^-$   & 1.78 & 1.85 $\pm$ 9.00E-2 \\

$K^+ / K^-$   & 1.78 &  1.80 $\pm$ 1.00E-1 \\



$ \Xi^+ /  \overline{ \Lambda} $   & 0.170 &  0.188 $\pm$ 3.90E-2  \\


$ ( \Xi^- +  \Xi^+ ) / ( \Lambda + \overline{\Lambda} ) $    & 0.115 & 0.130 $\pm$ 3.00E-2 \\

$ \Xi^+  / \Xi^-  $  &  0.193 & 0.232 $\pm$ 3.30E-2 \\

$ \Xi^+  / \Xi^-  $  & 0.193 &  0.247 $\pm$ 4.30E-2 \\

$ \Omega^+  / \Omega^-  $  & 0.357 &  0.383 $\pm$ 8.10E-2 \\

$ \Omega  / \Xi  $  &  0.153 &  0.219 $\pm$ 5.00E-2 \\

\hline
\end{tabular}
\caption{
SPS Pb+Pb at $\sqrt{s}$= 17 GeV, most central events. \protect\newline
Predicted versus experimental particle ratios for the best fit of our model.
%
}
\label{ratios_pbpb}
\end{table}

\begin{table}[ht]
\begin{tabular}{|lllllll|}
%
%
\hline
$\mu_b$   &  $\mu_s$     &  T   & $\lambda_s$ &  
$\rho_e$  &  $T_{eq, \rho_e}$   & $\lambda_s({eq, \rho_e})$   \\
GeV       &  GeV         &  GeV &             &
GeV/fm$^3$ & GeV               &     \\
\hline

0.239     &   0.0517    &  0.162  & 0.565 & 
0.408     &   0.169     &  0.422 \\


          &             &    +0.010    &   & 
	  &  +0.008     &    +0.025     \\

          &             &   -0.028   &    &
	  &  -0.030     &     -0.118    \\


\hline
  &    &         &  $\lambda_s$ &
  $\rho_n$  & $T_{eq, \rho_n}$  &  $\lambda_s({eq, \rho_n})$   \\
       &           &   &             &
       1/fm$^3$ & GeV               &     \\
\hline
%


     &       &   &     &  0.444   &   0.167          & 0.418 \\
     &       &   &     &          & +0.008           & +0.025 \\
     &       &   &     &          & -0.030           &  -0.117 \\

\hline
  &   &         &  $\lambda_s$ &  
$\rho_s$  & $T_{eq, \rho_s}$  &   $\lambda_s({eq, \rho_s})$    \\
       &           &   &             &
       1/fm$^3$ & GeV               &     \\
\hline


     &      &    &  & 2.8049   & 0.168   &  0.420  \\
     &      &    &  &    &  +0.009   & +0.026    \\
     &      &    &  &    &  -0.031   & -0.116    \\


\hline
\end{tabular}
\caption{
SPS Pb+Pb at $\sqrt{s}$= 17 GeV, most central events. \protect\newline
Thermodynamic parameters for the best fit and temperatures
and $\lambda_{\ s}$
extra\-polated to zero fugacities.
%
}
\label{t_pbpb}
\end{table}

\clearpage


\vspace{0.5cm}

\noindent{ \bf Central S+A collisions at 200 A GeV} 
\vspace{0.5cm}

\noindent
In table \ref{t_sau} we give the same quantities for central S+A collisions
at 200 A GeV with A=Au,W,Pb.
The data used are taken from reference \cite{nuclth9508020}.
Here we mainly used the $\overline{p}/p$ and $K^+/ K^-$ ratios to find
(T, $\mu_b$, $\mu_s$) 
while we also suppose exact strangeness conservation.
The resulted predicted ratios are 
$\overline{p}/p$ = 0.131 and
$K^+/ K^-$ = 1.59 to be compared with the experimental data 
$\overline{p}/p$ = 0.12 $\pm$ 0.02 and 
$K^+/ K^-$ = 1.59 $\pm$ 0.15.
The resulting temperature is  T=166 +10 -28 MeV.
The errors are taken to be percentually the same as in the Pb+Pb system.
We do not perform a full fit, as the resulting temperature 
is within the errors in agreement with reference
\cite{nuclth9508020}  where a full fit to many ratios is performed.
The calculated 
values in \cite{nuclth9508020} are T=165 $\pm$ 5 MeV, $\mu_b$=175 $\pm$ 5 MeV,
$\mu_s$=42.5 $\pm$ 4.5 MeV.

\begin{table}[ht]
\begin{tabular}{|lllllll|}
\hline
%
%
$\mu_b$   &  $\mu_s$     &  T   & $\lambda_s$ &  
$\rho_e$  &  $T_{eq, \rho_e}$   & $\lambda_s({eq, \rho_e})$   \\
GeV       &  GeV         &  GeV &             &
GeV/fm$^3$ & GeV               &     \\
\hline


0.191     &   0.0426   &  0.166   & 0.544 
& 0.424     &   0.169    &  0.422  \\


    &      &   +0.010   &  
&      &   +0.08 -0.030   &   +0.025 -0.118 \\

    &      &   -0.028  &  
&      &   +0.08 -0.030   &   +0.025 -0.118 \\

\hline
   &   &         &  $\lambda_s$ &
  $\rho_n$  & $T_{eq, \rho_n}$  &  $\lambda_s({eq, \rho_n})$   \\
	 &           &   &             &
	 1/fm$^3$ & GeV               &     \\
\hline
%


     &      &    &  
&  0.464  &   0.169  & 0.422  \\

     &      &    &  
&    &   +0.08 -0.030 &  +0.025 -0.118 \\

\hline
   &   &         &  $\lambda_s$ &  
$\rho_s$  & $T_{eq, \rho_s}$  &   $\lambda_s({eq, \rho_s})$    \\
       &           &   &             &
       1/fm$^3$ & GeV               &     \\
       \hline


    &     &    & 
& 2.92   & 0.169  &  0.422   \\

    &     &    & 
&    &  +0.08 -0.030  &   +0.025 -0.118  \\

\hline
\end{tabular}
\caption{
SPS S+A at $\sqrt{s}$= 19 GeV, most central events. \protect\newline
Thermodynamic parameters for the best fit and temperatures
and $\lambda_{\ s}$
extra\-polated to zero fugacities.
%
}
\label{t_sau}
\end{table}

\clearpage


\vspace{0.5cm}
\noindent{ \bf Central S+S collisions at 200 A GeV} \\
\vspace{0.5cm}

\noindent
In table \ref{t_ss} the results for the central S+S collisions at 200 A GeV
are shown, data are taken from references \cite{na35_papers}.
We use $K^0_s/ \Lambda$, 
$(B- \overline{B})/ h^-$,  $\overline{\Lambda} / \Lambda$,
$K^0_s/ h^-$ and $K^+ / K^-$ as well as strange\-ness con\-servation.
The resulting 
$\chi^2$/DOF =1.95/3=0.65 (CL $\sim$ 57\%)
imposing strangeness conservation is at T=182 +19 -8 MeV.
The minimum of $\chi^2$ ($\chi^2/DOF$=1.6/3=0.53 CL 66\%)
is at T=186 $\pm$ 12 MeV.

\begin{table}[ht]
\begin{tabular}{|lllllll|}
\hline

$\mu_b$   &  $\mu_s$     &  T   & $\lambda_s$ &  
$\rho_e$  &  $T_{eq, \rho_e}$   & $\lambda_s({eq, \rho_e})$   \\
GeV       &  GeV         &  GeV &             &
GeV/fm$^3$ & GeV               &     \\
\hline

0.218     &   0.058         &  0.182         &  0.654
& 0.182   &   0.187         &  0.473 \\


	  &                 & +0.019  & +0.065
&	  & +0.021    & +0.047       \\

	  &                 & -0.008  & -0.003
&	  &  -0.009   & -0.023      \\

  &   &    &   &
  $\rho_n$  & $T_{eq, \rho_n}$  &  $\lambda_s({eq, \rho_n})$   \\
	 &           &   &             &
	 1/fm$^3$ & GeV               &     \\
\hline

     &    &    & 
&  0.803  &   0.186   & 0.470 \\


     &    &    & 
&         & +0.009    &  +0.047  \\

     &    &    & 
&         &  -0.020   &  -0.023 \\

  &   &    &   &
$\rho_s$  & $T_{eq, \rho_s}$  &   $\lambda_s({eq, \rho_s})$    \\
       &           &   &             &
       1/fm$^3$ & GeV               &     \\
\hline

     &    &    & 
&  4.98  & 0.187   &  0.473  \\


     &    &    & 
&    &  +0.021   & +0.048    \\

     &    &    & 
&    &   -0.009  &  -0.023   \\

\hline
\end{tabular}
\caption{
SPS S+S at $\sqrt{s}$= 19 GeV, most central events. \protect\newline
Thermodynamic parameters for the best fit and temperatures
and $\lambda_{\ s}$
extra\-polated to zero fugacities.
%
}
\label{t_ss}
\end{table}

\clearpage

\vspace{0.5cm}
\vspace{0.5cm}

\subsubsection{Central nucleus nucleus collisions at $\sim$ 10 A GeV}

\noindent
Here we show results from central Si+Au collisions at 14.6 A GeV,
and from Au+Au collisions at 11.6 A GeV.
\vspace*{0.3cm}


\vspace{0.5cm}
\noindent{ \bf Central Si+Au collisions at 14.6 A GeV}
\vspace{0.5cm}

\noindent
In table \ref{t_siau} the results for central Si+Au collisions
at 14.6 A GeV are shown. 
We use the ratios $K/\pi$, $K^+/K^-$, $\overline{ \Lambda} / \Lambda$ and
$\phi/\pi$ from \cite{revions1}
as well as strangeness conservation.
We did not use 
 $\overline{p}$ in view of the large annihilation
expected at low energy.
The resulting $\chi^2/DOF$=6.8/2=3.4 (CL $\sim$ 3.3\%)
imposing strangeness conservation is at T=120 +8 -35 MeV.
We note however that the best $\chi^2$ ignoring strangeness conservation \\
($\chi^2/DOF$=0.092/2 with CL 63\%) is at $\mu_b/T$=4.05, $\mu_s/T$=0.9 
and T=102 MeV.

\begin{table}[ht]
\begin{tabular}{|lllllll|}
\hline



$\mu_b$   &  $\mu_s$     &  T   & $\lambda_s$ &  
$\rho_e$  &  $T_{eq, \rho_e}$   & $\lambda_s({eq, \rho_e})$   \\
GeV       &  GeV         &  GeV &             &
GeV/fm$^3$ & GeV               &     \\
\hline


0.540     &   0.107   &  0.120  & 0.548  &
0.170     &  147.4    & 0.347  \\

          &           &  +0.008  &  & 
          & +0.013    &   +0.052  \\

          &           &   -0.035  & & 
          &  -0.056   &    -0.250 \\

\hline
$\mu_b$   &  $\mu_s$  &  T       &  $\lambda_s$ &
  $\rho_n$  & $T_{eq, \rho_n}$  &  $\lambda_s({eq, \rho_n})$   \\
	 &           &   &             &
	 1/fm$^3$ & GeV               &     \\
\hline


  &  &   &  &   
 0.182 &  0.141  &  0.322   \\

  &  &   &  &   
  &  +0.013  &  +0.048   \\

  &  &   &  &   
  &  -0.053 &  -0.23  \\

\hline
$\mu_b$   &  $\mu_s$  &  T       &  $\lambda_s$ &  
$\rho_s$  & $T_{eq, \rho_s}$  &   $\lambda_s({eq, \rho_s})$    \\
       &           &   &             &
       1/fm$^3$ & GeV               &     \\
\hline


  &  &   &  &   
1.161  &  0.143  &  0.330   \\

  &  &   &  &   
  &   +0.013  &   +0.049   \\

  &  &   &  &   
  &    -0.054 &    -0.237  \\

\hline

\end{tabular}
\caption{
AGS Si+Au at $\sqrt{s}$= 5.4 GeV, most central events. \protect\newline
Thermodynamic parameters for the best fit and temperatures
and $\lambda_{\ s}$
extra\-polated to zero fugacities.
%
}
\label{t_siau}
\end{table}

\clearpage


\vspace{0.5cm}
\noindent{\bf Central Au+Au collisions at 11.6 A GeV}
\vspace{0.5cm}

\noindent
The results for central Au+Au collisions at 11.6 A GeV
are shown in tables \ref{ratios_auau}, \ref{t_auau}.
We used $K^+/K^-$, $K/ \pi$, $p/\pi^+$ and $K/\Lambda$ ratios from
\cite{revions1}.
The resulting $\chi^2/DOF$=0.99/2=0.496 (CL $\sim$ 60\%)
imposing strangeness conservation is at T=96 +4 -5 MeV.
The minimum of $\chi^2$ ($\chi^2/DOF$=0.92/2=0.46 CL 63\%) is at T=95
MeV.

\begin{table}[ht]
\begin{tabular}{|lll|}
\hline
ratio  & model & data  \\
\hline

$<K>/ \Lambda$  & 0.562  &  0.675 $\pm$ 0.144  \\

$K^+ / K^-$   &   5.65  & 6.303 $\pm$ 2.55 \\

$p/ \pi^+$   &  1.088  &  1.098 $\pm$ 0.127  \\

$<K>/ <\pi>$    &  7.299E-2  &   8.080E-2 $\pm$ 1.420E-2  \\

\hline
\end{tabular}
\caption{ 
AGS Au+Au at $\sqrt{s}$= 4.9 GeV, most central events. \protect\newline
Predicted versus experimental particle ratios for the best fit of our model.
%
}
\label{ratios_auau}
\end{table}

\begin{table}[ht]
\begin{tabular}{|lllllll|}
\hline
%
$\mu_b$   &  $\mu_s$     &  T   & $\lambda_s$ &  
$\rho_e$  &  $T_{eq, \rho_e}$   & $\lambda_s({eq, \rho_e})$   \\
GeV       &  GeV         &  GeV &             &
GeV/fm$^3$ & GeV               &     \\
\hline

0.563     &   0.084   &  0.096   & 0.281
& 0.484E-01    &   0.121  &  0.233  \\

    &     &   $\pm$ 0.005  & 
&   &    $\pm$ 0.08  &  $\pm$ 0.037 \\

\hline
  &   &    &   &
  $\rho_n$  & $T_{eq, \rho_n}$  &  $\lambda_s({eq, \rho_n})$   \\
	 &           &   &             &
	 1/fm$^3$ & GeV               &     \\
\hline

  &    &    & 
&  0.602E-01  &   0.113   & 0.196  \\

  &    &    & 
&   &    $\pm$ 0.079  & $\pm$ 0.037 \\

\hline
  &   &    &   &
$\rho_s$  & $T_{eq, \rho_s}$  &   $\lambda_s({eq, \rho_s})$    \\
       &           &   &             &
       1/fm$^3$ & GeV               &     \\
\hline

  &    &    & 
&  0.387  &  0.116  &  0.210  \\

  &    &    & 
&   &  $\pm$ 0.08 &   $\pm$ 0.037 \\

\hline
\end{tabular}
\caption{
AGS Au+Au at $\sqrt{s}$= 4.9 GeV, most central events. \protect\newline
Thermodynamic parameters for the best fit and temperatures
and $\lambda_{\ s}$
extra\-polated to zero fugacities.
}
\label{t_auau}
\end{table}

\clearpage

\vspace{0.5cm}
\vspace{0.5cm}

\subsubsection{Central nucleus nucleus collisions at $\sim$ 2 A GeV}


\vspace{0.5cm}
\noindent{ \bf Central Ni+Ni collisions at 1.9 A GeV}
\vspace{0.5cm}

\noindent
The results for central Ni+Ni collisions at 1.8 A GeV 
are shown in tables \ref{ratios_nini}, \ref{t_nini}.
We used the measured $K^+/K^-$ and  $K/\pi$ ratios
 from \cite{fopi,senger_stroebele} (figure 4.4)
 and used additionally the $K/\Lambda$ ratio
 deduced as $\Lambda \sim  \Lambda - \overline{\Lambda}  =
 2( K^+ - K^-)$ and imposed strangeness conservation.
 We add a systematic error of 15\% for the feeding experimental uncertainty
 linearly.
The resulting $\chi^2/DOF$=2.16/1 (CL $\sim$ 10\%)
imposing strangeness conservation is at T=0.044 +0.0023  -0.0002 GeV.
The minimum of $\chi^2$ ($\chi^2/DOF$=0.025/1 with CL better than 90\%) is at 
T=0.0448 GeV.

\begin{table}[ht]
\begin{tabular}{|lll|}
\hline
ratio  & model & data  \\
\hline

$<K>/ \Lambda$  & 0.373 & 0.275 $\pm$ 6.72E-2  \\

$K^+ / K^-$  & 20.90 & 20.86 $\pm$ 6.88  \\

$<K>/ <\pi>$  &  3.16E-3  &  3.20E-3 $\pm$ 9.10E-4  \\

\hline
\end{tabular}
\caption{ 
GSI Ni+Ni at $\sqrt{s}$= 2.6 GeV, most central events. \protect\newline
Predicted versus experimental particle ratios for the best fit of our model.
%
}
\label{ratios_nini}
\end{table}

\begin{table}[ht]
\begin{tabular}{|lllllll|}
\hline

$\mu_b$   &  $\mu_s$     &  T   & $\lambda_s$ &  
$\rho_e$  &  $T_{eq, \rho_e}$   & $\lambda_s({eq, \rho_e})$   \\
GeV       &  GeV         &  GeV &             &
GeV/fm$^3$ & GeV               &     \\
\hline


0.678     &   0.102   &  0.044  & 8.93E-3 
&   0.987E-03  &   0.602E-01  &   1.55E-02 \\


    &      &  $\pm$ 0.0023    &   $\pm$ 0.00335
&     &   $\pm$ 0.0114   &   +0.0435  \\

    &      &      &   
&     &      &    -5.55E-3  \\

\hline
  &   &    &   &
  $\rho_n$  & $T_{eq, \rho_n}$  &  $\lambda_s({eq, \rho_n})$   \\
	 &           &   &             &
	 1/fm$^3$ & GeV               &     \\
\hline


  &    &    & 
& 0.161E-02  & 0.504E-01    &  5.16E-03   \\


  &    &    & 
&   &  $\pm$ 0.0096      &  +0.0144  \\

  &    &    & 
&   &        &  -1.84E-3  \\

\hline
  &   &    &   &
$\rho_s$  & $T_{eq, \rho_s}$  &   $\lambda_s({eq, \rho_s})$    \\
       &           &   &             &
       1/fm$^3$ & GeV               &     \\
\hline


  &    &    & 
& 1.18E-02   & 5.32E-2   &  7.38E-03   \\


  &    &    & 
&  &  $\pm$ 0.0104   &  +0.0207     \\

  &    &    & 
&  &     &   -2.64E-3    \\

\hline
\end{tabular}
\caption{
GSI Ni+Ni at $\sqrt{s}$= 2.6 GeV, most central events. \protect\newline
Thermodynamic parameters for the best fit and temperatures
and $\lambda_{\ s}$
extra\-polated to zero fugacities.
%
}
\label{t_nini}
\end{table}

\clearpage

\vspace{0.5cm}
\vspace{0.5cm}

\subsection{Electron positron reactions}


\noindent
The primary $q \overline{q}$ production in $e^+ e^-$ collisions 
is  not thermal due to the hard nature of the primary $\gamma$, $Z$ couplings.
However as long as the final particle 
multiplicity is much higher than 2, it is conceivable that this fact does 
not matter and the subsequent fragmentation of quark-antiquark jets
into hadrons is thermal.
We used the $K/\pi$, $\rho/\pi$, $\pi/p$ and $\Delta/\pi$ ratios.

\noindent
Our results for $e^+ e^-$ collisions at $\sqrt{s}$=91 GeV are
shown in table \ref{t_e+e-_91gev}.
We used the data from reference \cite{Chliapnikov} in particular the
initially produced particles from the Pei model (table 1 in \cite{Chliapnikov}).
We used as systematic error of the feeding correction
performed in this model, the difference between the Pei and the Jetset model
(also in table 1 in \cite{Chliapnikov}).
 The resulting minimum $\chi^2/DOF$=2.52/3=0.84 (CL $\sim$ 47\%)
 is at T=145 +39 -45 MeV and  $\lambda_s$ is 0.338 +0.127  -0.2.

\noindent
In reference \cite{becattini} a temperature of 160.6 $\pm$1.7 $\pm$ 3.1
is given for $e^+e^-$ collisions at 91 GeV, however with a not
acceptable $\chi^2/DOF$ of 60.8/21, having a CL of $10^{-5}$.
In this reference strangeness has been weigted by a factor
$\gamma_s$ which is a free parameter of the fit.
In reference \cite{Chliapnikov} a temperature of T=142.4 $\pm$ 0.018 MeV and
$\lambda_s$=0.295 $\pm$ 0.006
are extracted in $e^+e^-$ collisions at 91 GeV.
Within our large errors we agree with both reference \cite{becattini} and
\cite{Chliapnikov}.
Our best value of the temperature is however nearer to
 the results of reference \cite{Chliapnikov}.

\begin{table}[ht]
\begin{tabular}{|lllllll|}
\hline
$\mu_b$   &  $\mu_s$     &  T   & $\lambda_s$ &  
$\rho_e$  &  $\rho_n$ &  $\rho_s$   \\
GeV       &  GeV         &  GeV &         &
GeV/fm$^3$ & 1/fm$^3$    & 1/fm$^3$  \\
\hline

0.     &   0.   &  0.145          & 0.338             & 0.152    &   0.208   &  1.25 \\
       &        &  +0.039 -0.045  & +0.127 -0.20      &     &     &  \\

\hline
\end{tabular}
\caption{
$e^+$+ $e^-$ collisions at $\sqrt{s}$= 91 GeV. \protect\newline
Thermodynamic parameters for the best fit and temperatures
and $\lambda_{\ s}$
extra\-polated to zero fugacities.
}
\label{t_e+e-_91gev}
\end{table}

\clearpage

\vspace{0.5cm}
\vspace{0.5cm}

\subsection{Hadronic reactions}

\subsubsection{Proton proton reactions}

\noindent
Here we show results from
p+p collisions at $\sqrt{s}$=17 and 27 GeV.
\vspace{0.3cm}


\vspace{0.5cm}
\noindent{ \bf Proton proton reactions at $\sqrt{s}$=27 GeV}
\vspace{0.5cm}

\noindent
For p+p collisions at $\sqrt{s}$=27 GeV 
we used  9 measured ratios from reference \cite{becattini} namely
$K/\pi$, $\overline{p}/p$, $\pi/p$, $\eta/\pi^0$, $\Lambda/ K^0_s$,
$K^+/K^-$, $\overline{ \Lambda } / \Lambda$,
$\phi/ \pi$, $\Delta^{++} / p$  and imposed strangeness conservation.
We add a systematic error of 15\% linearly to account for the
experimental feeding uncertainties.
The predicted and 
the experimental ratios are shown in table \ref{ratios_pp27gev}.
We find a temperature of 128 +5 -9 MeV, with
a $\chi^2/DOF$=71/5=14.2, which  has a CL of the order $10^{-14}$,
therefore the fit is not acceptable (table \ref{t_pp27gev}).
\vspace{0.3cm}

\noindent
In reference \cite{hepph9702274} the authors discuss p+p collisions
at 4 different $\sqrt{s}$.
For $\sqrt{s}$=27 GeV considered here,
the obtained $\chi^2/DOF$ is 136.4/27 with a CL much less than $10^{-5}$.
The obtained temperature in \cite{hepph9702274} is 169 $\pm$ 2.1 $\pm$ 3.4.
\vspace{0.3cm}

\noindent
The bad quality of the fits of p+p collisions at 
$\sqrt{s}$=27 GeV renders the resulting temperatures questionable.
Nevertheless we point out that the two analyses 
yield incompatible temperature values
within the errors.
The analysis of \cite{hepph9702274} does not introduce systematic errors as we do.
However, it allows for an arbitrary weighting factor (called $\gamma_s$)
acting only on strange particles.
This factor is similar to an
increase of the systematic 
experimental errors of strange particle multiplicities only.

\begin{table}[ht]
\begin{tabular}{|lll|}
\hline
ratio  & model & data  \\
\hline

$K / \pi$    & 0.120 & 7.35E-2 $\pm$ 1.80E-2 \\

$\overline{p}/p$   & 2.65E-2 &  5.25E-2 $\pm$  1.36E-2 \\

$\pi / p$  & 5.51  & 3.14 $\pm$ 0.846 \\

$eta/ \pi^0$  & 6.89E-2 &  7.75E-2  $\pm$ 3.15E-2  \\

$\Lambda / K^0_s$  & 0.533 &  0.539  $\pm$ 0.158 \\

$K^+ / K^- $  & 1.52 &  1.48 $\pm$ 0.364 \\

$\overline{ \Lambda }/ \Lambda $  &  3.88E-2 & 0.160 $\pm$ 6.75E-2 \\

$\phi/ \pi$  & 1.07E-2 & 5.034E-3 $\pm$ 1.34E-3 \\

$\Delta^{++} / p$  & 0.148 &  0.182 $\pm$ 4.97E-2  \\

\hline

\end{tabular}
\caption{
p+p collisions at $\sqrt{s}$= 27 GeV. \protect\newline
Predicted versus experimental particle ratios for the best fit of our model.
The $\chi^2/DOF$ of this fit is not acceptable.
}
\label{ratios_pp27gev}
\end{table}

\begin{table}[ht]
\begin{tabular}{|lllllll|}
\hline

$\mu_b$  (GeV) &  $\mu_s$  (GeV)   &  T  (GeV) & $\lambda_s$ &  
$\rho_e$  &  $T_{eq, \rho_e}$   & $\lambda_s({eq, \rho_e})$   \\
GeV       &  GeV         &  GeV &             &
GeV/fm$^3$ & GeV               &     \\
\hline


0.243     &   0.0282   &  0.128   & 0.329  
& 0.830E-1    &   0.132  &  0.283  \\


     &      &   +0.005   &   
&     &    +0.005  &   +0.0218  \\

     &      &    -0.009  &   
&     &     -0.010 &    -0.045 \\

\hline
  &   &    &   &
  $\rho_n$  & $T_{eq, \rho_n}$  &  $\lambda_s({eq, \rho_n})$   \\
	 &           &   &             &
	 1/fm$^3$ & GeV               &     \\
\hline


  &    &    & 
&  0.123  &  0.131   & 0.279  \\


  &    &    & 
&    &   +0.005   &  +0.0215   \\

  &    &    & 
&    &    -0.010  &    -0.044 \\

\hline
  &   &    &   &
$\rho_s$  & $T_{eq, \rho_s}$  &   $\lambda_s({eq, \rho_s})$    \\
       &           &   &             &
       1/fm$^3$ & GeV               &     \\
\hline


  &    &    & 
&  0.744  & 0.132    &  0.283  \\


  &    &    & 
&    &  +0.005    &  +0.0218  \\

  &    &    & 
&    &   -0.010   &   -0.045 \\

\hline
\end{tabular}
\caption{
p+p collisions at $\sqrt{s}$= 27 GeV. \protect\newline
Thermodynamic parameters for the best fit and temperatures
and $\lambda_{\ s}$
extra\-polated to zero fugacities.
The $\chi^2/DOF$ of this fit is not acceptable.
%
}
\label{t_pp27gev}
\end{table}

\clearpage

\vspace{0.5cm}
\noindent{ \bf Proton proton reactions at $\sqrt{s}$=17 GeV}
\vspace{0.5cm}

\noindent
We used 6 measured ratios from references \cite{siklerqm99,na49phi} namely
$K/\pi$, $\overline{p}/p$, $\pi/p$, $K^+/K^-$, 
$\phi/ \pi$, $K / p$  and imposed strangeness conservation.
We add a systematic error of 15\% linearly to account for the
experimental feeding uncertainties.
The resulting $\chi^2/DOF$ is (322/4)=80.6 (CL of the order $10^{-66}$ !)
for a temperature of 144 MeV, is completely unacceptable.
\vspace{0.3cm}

\noindent
The predicted ratios in table \ref{ratios_pp158gev_syst15}
are overestimating $K^+$, $K^-$ and $\phi$  yields, 
therefore suggesting 
that if thermal conditions prevail,
then at least  two thermal reservoires are present.
This may be due to the influence of diffractive processes
in minimum bias triggers.
\vspace{0.3cm}

\noindent
We therefore repeat the fit with only ratios which are not
much influenced by diffractive processes namely
 $K/\pi$, $K^+/K^-$, and $\phi/ \pi$.
Note that the temperature at the minimum of $\chi^2$
does not satisfy strangeness conservation.
This is combatible with the hypothesis of the importance of
diffractive processes showing up in leading baryons (e.g. p, $\Lambda$).
Imposing strangeness conservation the fit
results in a temperature of 144 MeV
with a bad $\chi^2/DOF$=169/1 (CL $\sim$ $10^{-39}$).
At the minimum of the $\chi^2$ the temperatures is 96 +8 -9 MeV
with a better $\chi^2/DOF$=5.13/1 (CL $\sim$ 2\%).
These results are shown in tables 
\ref{ratios_pp158gev}, \ref{t_pbpb158gev}.
\noindent
The very small errors resulting for p+p collisions at $\sqrt{s}$=17
and 27 GeV
should not be trusted because of the bad CL of the fit.
\vspace*{0.3cm}

\begin{table}[ht]
\begin{tabular}{|lll|}
\hline
ratio  & model & data  \\
\hline

$ K/\pi$   & 0.136 & 5.97E-2 $\pm$ 9.90E-3  \\

$\overline{p}/p$  & 5.97E-2 & 4.68E-2 $\pm$ 8.06E-3  \\

$\pi/p$  & 4.261 & 3.59 $\pm$ 0.558 \\

$K^+/K^-$  & 1.55 & 1.61 $\pm$ 0.277 \\

$\phi/\pi$   & 1.83  & 4.25E-3 $\pm$ 1.17E-3\\

$K/p$  & 0.578  & 0.214 $\pm$ 3.56E-2 \\

\hline

\end{tabular}
\caption{
p+p collisions at $\sqrt{s}$= 17 GeV. \protect\newline
Predicted versus experimental particle ratios for the worst fit of our model.
%
}
\label{ratios_pp158gev_syst15}
\end{table}

\begin{table}[ht]
\begin{tabular}{|lll|}
\hline
ratio  & model & data  \\
\hline

$ K/\pi$   & 0.0723 & 5.97E-2 $\pm$ 9.90E-3  \\

$\overline{p}/p$  & 5.36E-2 & 4.68E-2 $\pm$ 8.06E-3  \\

$\pi/p$  & 47.7 & 3.59 $\pm$ 0.558 \\

$K^+/K^-$  & 1.60 & 1.61 $\pm$ 0.277 \\

$\phi/\pi$   & 2.06E-3  & 4.25E-3 $\pm$ 1.17E-3\\

$K/p$  & 3.44  & 0.214 $\pm$ 3.56E-2 \\

\hline

\end{tabular}
\caption{
p+p collisions at $\sqrt{s}$= 17 GeV. \protect\newline
Predicted versus experimental particle ratios for the best fit of our model.
Only three ratios are fitted ($\phi/\pi$, $K^+/K^-$, $K/ \pi$).
}
\label{ratios_pp158gev}
\end{table}

\begin{table}[ht]
\begin{tabular}{|lllllll|}
\hline

$\mu_b$   &  $\mu_s$     &  T   & $\lambda_s$ &  
$\rho_e$  &  $T_{eq, \rho_e}$   & $\lambda_s({eq, \rho_e})$   \\
GeV       &  GeV         &  GeV &             &
GeV/fm$^3$ & GeV               &     \\
\hline


0.222    &   0.0277   &  0.096   &  0.151  & 0.0124    &   0.096   &  0.121  \\

     &      &   +.008    &     &                        &    +0.008 &   +0.034  \\

     &      &    -.009   &                  &  		&    -0.010 &  -0.039 \\

\hline
  &   &    &   &
  $\rho_n$  & $T_{eq, \rho_n}$  &  $\lambda_s({eq, \rho_n})$   \\
	 &           &   &             &
	 1/fm$^3$ & GeV               &     \\
\hline


  &    &    & 
&  0.0293  &   0.096   & 0.121  \\

  &    &    & 
&    &    +0.008 -0.010  &  +0.034 -0.039 \\

\hline
  &   &    &   &
$\rho_s$  & $T_{eq, \rho_s}$  &   $\lambda_s({eq, \rho_s})$    \\
       &           &   &             &
       1/fm$^3$ & GeV               &     \\
\hline
  &    &    & 
& 0.157  & 0.097    &  0.125  \\
  &    &    & 
&   &  +0.008 -0.010   &  +0.035 -0.04 \\
\hline
\end{tabular}
\caption{
p+p collisions at $\sqrt{s}$= 17 GeV. \protect\newline
Thermodynamic parameters for the best fit and temperatures
and $\lambda_{\ s}$
extra\-polated to zero fugacities.
Only three ratios are fitted ($\phi/\pi$, $K^+/K^-$, $K/ \pi$).
%
}
\label{t_pbpb158gev}
\end{table}

\clearpage

\vspace{0.5cm}
\vspace{0.5cm}
\subsubsection{Proton antiproton reactions}

\noindent
Here we show results from $p+ \overline{p}$ collisions at $\sqrt{s}$=900 GeV  and
central and peripheral
$p+ \overline{p}$ collisions at $\sqrt{s}$=1.8 TeV .
\vspace{0.5cm}

\vspace{0.5cm}
\noindent{\bf Central proton antiproton reactions at $\sqrt{s}$=900 GeV}
\vspace{0.5cm}

\noindent
For $p+ \overline{p}$ collisions at $\sqrt{s}$=900 GeV we used
 5 measured ratios out of the multiplicities from
  reference \cite{hepph9702274}.
  in particular
$K^0_s/charged$ and $N/charged $, $\Lambda / K^0_s$, $\Lambda / \Xi^-$,
$\Xi^-/charged$.
 The predicted and the experimental ratios are shown in table \ref{ratios_ppbar900gev}.
The temperature from the fit is:
\vspace*{0.3cm}

T($p+ \overline{p}$ $\sqrt{s}$=900 GeV) = 143 +21 -37 MeV 
\vspace*{0.3cm}

\noindent
with a $\chi^2/DOF$ of 4.48/4 having a CL of $\sim$ 34\%, while
$\lambda_s$ is
\vspace*{0.3cm}

$\lambda_s$($p+ \overline{p}$ $\sqrt{s}$=900 GeV) = 0.330 +0.077 -0.166.
\vspace*{0.3cm}

\noindent
We add linearly a systematic error of 15\% to account for experimental feeding 
uncertainties.
The experimental errors are very large (e.g. 57\% for $\Xi^-$)
therefore the resulting temperature has also a large error.
\vspace*{0.3cm}

\begin{table}[ht]
\begin{tabular}{|lll|}
\hline
ratio  & model & data  \\
\hline

$K^0_s/charged $   &  5.73E-2 &  3.85E-2 $\pm$ 1.06E-2 \\

$n/charged$  &  2.17E-2 &  2.81E-2 $\pm$ 1.25E-2  \\

$\Lambda / K^0_s$  & 0.181 &  0.277 $\pm$ 0.114 \\

$\Lambda / \Xi^-$  & 7.96 & 10.86 $\pm$ 8.344 \\

$\Xi^-/charged$   & 1.305E-3 & 9.83E-4 $\pm$ 7.12E-4 \\

\hline

\end{tabular}
\caption{ 
p+$\overline{p}$ collisions at $\sqrt{s}$= 900 GeV. \protect\newline
Predicted versus experimental particle ratios for the best fit of our model.
%
}
\label{ratios_ppbar900gev}
\end{table}

\clearpage

\vspace{0.5cm}
\noindent{ \bf Peripheral proton antiproton reactions at $\sqrt{s}$=1.8 GeV}
\vspace{0.5cm}

\noindent
For  the most peripheral $p+ \overline{p}$ collisions at $\sqrt{s}$=1.8 TeV 
we used  2 measured ratios from reference \cite{Gutay} namely
$K/\pi$ and $\overline{p}/ \pi $ (table \ref{ratios_periph_ppbar18tev}).
We use the only  2 available ratios from experiment, which with
zero fugacities and temperature as the only parameter
leaves one degree of freedom.
The most peripheral collisions are defined as the
ones with the lowest measured charge multiplicity.
The temperature from the fit is 
\vspace*{0.3cm}

T(peripheral $p+ \overline{p}$ $\sqrt{s}$=1.8 Te)=
140 $\pm$ 8 MeV 
\vspace*{0.3cm}

\noindent
with a $\chi^2/DOF$ of 18.04/1 having  a CL of $\sim$ 2.0E-4,
therefore the fit is very bad.
We add a systematic error of 0.004 linearly to account for the
experimental feeding uncertainties.
This error has been estimated from deviations between the ratios found in
different experimental runs with $p+ \overline{p}$ collisions at $\sqrt{s}$=1.8 TeV
\cite{phd_zhan}.
 The predicted and the experimental ratios are shown in table \ref{ratios_periph_ppbar18tev}.
Then the resulting temperature is 154 $\pm$ 9 MeV
and the $\chi^2/DOF$ is 6.36/1 having  a CL of $\sim$ 1.1\%.
The quality of the fit is low as in the minimum bias p+p data at lower energies.
$\lambda_s$ is
\vspace*{0.3cm}

$\lambda_s$(peripheral $p+ \overline{p}$ $\sqrt{s}$=1.8 Te)=
0.372 +0.031 -0.034.
\vspace*{0.3cm}

\begin{table}[ht]
\begin{tabular}{|lll|}
\hline
ratio  & model & data  \\
\hline

$K/ \pi$   & 0.144  & 0.115 $\pm$ 1.20E-2 \\

$\overline{p}/ \pi$  & 6.87E-2 & 7.40E-2 $\pm$ 9.00E-3 \\

\hline
\end{tabular}
\caption{  
Most peripheral p+$\overline{p}$ collisions at $\sqrt{s}$= 1.8 TeV. \protect\newline
Predicted versus experimental particle ratios for the best fit of our model.
}
\label{ratios_periph_ppbar18tev}
\end{table}

\clearpage

\vspace{0.5cm}
\noindent{\bf Central proton antiproton reactions at $\sqrt{s}$=1.8 TeV}
\vspace{0.5cm}

\noindent
The results for the most central $p+ \overline{p}$ collisions 
at $\sqrt{s}$=1.8 TeV are shown in table \ref{ratios_ppbar18tev}.
The most central collisions are defined as the
ones with the highest measured charge multiplicity.
For  the most central  $p+ \overline{p}$ collisions at $\sqrt{s}$=1.8 TeV 
we used again the 2 measured ratios from reference \cite{Gutay} namely
$K/\pi$ and $\overline{p}/ \pi $.
The predicted 
and the experimental ratios are shown in table \ref{ratios_ppbar18tev}.
The temperature from the fit is 
\vspace*{0.3cm}

T(central $p+ \overline{p}$ $\sqrt{s}$=1.8 TeV)= 150 $\pm$ 9 MeV 
\vspace*{0.3cm}

\noindent
with a $\chi^2/DOF$ of 2.28E-3/1 having a CL of $\sim$ 96\%.
$\lambda_s$ is 
\vspace*{0.3cm}

$\lambda_s$(central $p+ \overline{p}$ $\sqrt{s}$=1.8 TeV)=
0.357 +0.035 -0.033.
\vspace*{0.3cm}

\noindent
We did not add a systematic error.
\vspace*{0.3cm}

\begin{table}[ht]
\begin{tabular}{|lll|}
\hline
ratio  & model & data  \\
\hline

$K/ \pi$   & 0.1424  & 0.1420 $\pm$ 2.00e-2  \\

$\overline{p}/ \pi$  & 6.24e-2  & 6.20e-2 $\pm$ 9.00E-3 \\

\hline
\end{tabular}
\caption{  
Most central p+$\overline{p}$ collisions at $\sqrt{s}$= 1.8 TeV. \protect\newline
Predicted versus experimental particle ratios for the best fit of our model.
}
\label{ratios_ppbar18tev}
\end{table}

\noindent
We think that the dramatic change in the quality of the fit
between peripheral and central $p + \overline{p}$ collisions can not be
explained by overestimated experimental errors in the central collisions 
respectively by underestimated experimental errors in the  peripheral collisions
alone.
Barring 
 the obvious possibility that peripheral $p + \overline{p}$ collisions are non thermal
the situation
 supports our previous conjecture on the important role
of diffractive processes in peripheral and minimum bias
hadronic collisions.
\vspace*{0.3cm} 

\begin{table}[ht]
\begin{tabular}{|lll|}
\hline
   collision   &  $\chi^2/DOF$    (CL)  &  $\chi^2/DOF$  (CL) no SC \\
\hline

Au+Au $\sqrt{s}$=130 GeV   & 1.41/1 (CL 23\%)  &        \\ 
with syst. error 15\%   & 0.27/1 (CL 60\%)  & \\
\hline

Pb+Pb $\sqrt{s}$=17 GeV   & 16.1/12=1.34 (CL 20\%)    & 11.9/11=1.08 (CL 50\%)  \\ 
\hline

 S+A $\sqrt{s}$=19 GeV &   no fit performed     &             \\ 
\hline

S+S $\sqrt{s}$=19 GeV   & 1.95/3=0.65 (CL 57\%) &  1.6/3=0.53 (CL 66\%)   \\ 
\hline

Si+Au $\sqrt{s}$=5.4 GeV   & 6.8/2=3.4 (CL 3.3\%)  & 0.092/2=0.046     \\ 
  & & (CL 63\%) \\
\hline

Au+Au $\sqrt{s}$=4.9 GeV   & 0.99/2=0.50 (CL 60\%)    &   0.92/2=0.46 (CL 63\%)   \\ 
\hline

Ni+Ni $\sqrt{s}$=2.6 GeV   & 2.16/1 (CL 10\%)  & 0.025/1 (CL 90\%)   \\ 
\hline

p+p $\sqrt{s}$=17 GeV   & 169/1 (CL $10^{-39}$)  &  5.13/1   (CL 2\%)  \\ 
min. bias (3 ratios) & & \\
\hline

p+p $\sqrt{s}$=27 GeV   
&  71/5=14.2 (CL $10^{-14}$)     & 68.6/5=13.7  (CL $10^{-13}$) \\ 
min. bias & & \\
\hline

$p + \overline{p}$ $\sqrt{s}$=900 GeV  & 4.48/4  (CL 34\%) &      \\ 
min. bias & & \\
\hline

$p + \overline{p}$ $\sqrt{s}$=1.8 TeV  & 2.3E-3/1 (CL 96\%)   &     \\ 
\hline

$p + \overline{p}$ $\sqrt{s}$=1.8 TeV   &   &   \\ 
peripheral      &  18/1 (CL 2E-4)   &   \\
no syst. error & & \\
periph.    &  6.36/1 (CL 1.1\%)   &   \\ 
with syst. error & & \\
\hline

$e^+ e^-$  $\sqrt{s}$=91 GeV & 2.52/3=0.84 (CL 47\%) &     \\ 
minimum bias  & & \\
\hline

\end{tabular}
\caption{Chisquare per degree of freedom and confidence level (CL) for each
collision system analysed.
The reactions are taken with central trigger unless differently stated. Strangeness
conservation is taken into account in the second row while it is not in the 3d row
(denoted by SC=strangeness conservation).}
\label{chisquare}
\end{table}

\clearpage

\vspace{0.5cm}
\vspace{0.5cm}

\subsection{The initial energy density estimation}

\noindent
To estimate the initial energy density achieved in each collision we use several
methods. The first method $(\alpha)$ is based on Bjorken's  formula
\cite{bjorkenform}:
\begin{equation}
\epsilon_{\ in} = \frac { (dE_T / d y)_{ycm} }  { \pi R_T^2 \tau }
\label{bjorken}
\end{equation}

\noindent
where $(dE_T / d y)_{ycm}$  is the transverse energy ($E_T=\sqrt{ p_T^2 + m^2}$) per
unit rapidity at midrapidity, $R_T$ is the transverse radius of the particle source
after a formation time $\tau$ taken to be 1 fm/c.

\noindent
We estimate the $(dE_T / d y)_{ycm}$ when not available from experiment, as follows.
We first estimate the total $E_T$ as

\begin{equation}
(dE_T/dy)^{tot}(ycm) = \sum_{\alpha} \ [ \ (dN/dy)_{\alpha}(ycm) 
\sqrt{ \left \langle p_T \right \rangle_{\alpha}^2 + m_{\alpha}^2 } 
\ ]
\label{et_tot}
\end{equation}

\noindent
(Method ($\alpha$))
\vspace{0.3cm}

\noindent
where the sum runs over particles of type $\alpha$ with multiplicity per
rapidity unit at midrapidity (ycm) $(dN/dy)_{\alpha}$,
mean transverse momentum $\left \langle p_T \right \rangle_{\alpha}$
and mass $m_{\alpha}$.
The mean transverse momentum is also taken at midrapidity, when
available.
We take three types of particles: $\pi$, $K$ and nucleons, as well as
 antinucleons when available.
To calculate the 
differencial $(dN / d y)_{ycm}$, when not available,
we first find the mean $(dN / d y)$,  dividing the total particle
multiplicity by the total
rapidity interval and then we use an extrapolation factor
A to extract the value at midrapidity.
This factor is found from measured rapidity distributions
of particles at or as near as possible to the $\sqrt{s}$ considered.
This may vary from nuclear to hadronic reaction and with $\sqrt{s}$.
In the cases where the $dN/dy$ of the particles are measured at midrapidty
we use  these values.
\vspace{0.3cm}

\noindent
The formation time is taken in the literature usually in all reactions 1 fm/c,
and this is what we use.
In the following we give some examples of our $\epsilon_{\ in}$ estimation
with method ($\alpha$).
\vspace*{0.3cm} 
\vspace{0.3cm}

\noindent
\underline{\bf 1. $e^+ + e^-$ } \hspace{0.2cm}
For the $e^+ e^-$ collisions at $\sqrt{s}$=91 GeV we used the multiplicities from 
\cite{Chliapnikov}.
For two jet events in  $e^+e^-$ collisions 
the jet axis is the longitudinal axis defining rapidity and $p_T$ refers to it.
We used the $dN/dy$ at midrapidity of $\pi$, $K$, $p$  in quark jets
produced in $e^+ e^-$ collisions at $\sqrt{s}$=91 GeV by the DELPHI
collaboration \cite{delphi}.

We used the mean transverse momenta from hadron hadron collisions
at high $\sqrt{s}$.
In particular we use the mean transverse momenta from $p+ \overline{p}$
collisions at 1.8 TeV from \cite{Gutay} at charged multiplicity N$\sim45$ (to represent
minimum bias values) namely
\vspace{0.3cm}

$\left \langle p_T \right \rangle_{\pi}$= 0.34 GeV,
\vspace{0.3cm}

$\left \langle p_T \right \rangle_{K}$= 0.5 GeV and
\vspace{0.3cm}

$\left \langle p_T \right \rangle_{p}$= 0.59 GeV.
\vspace{0.3cm}

\noindent
In hadron collisions the mean transverse momenta do not change much with
  $\sqrt{s}$ above 10 GeV (see e.g. figure 2.7 in \cite{senger_stroebele}).
The transverse size of the 
initial hadronic system of quarks and antiquarks
produced, is given by the uncertainty principle, as 1 over the average $p_T$.
For the transverse radius of $e^+e^-$ collisions we take therefore
the inverse
mean $p_T$ of 0.34 GeV, giving $R_T$=0.6 fm.
The  resulting initial energy density is:
\vspace{0.3cm}

$\epsilon_{\ in}$($e^+e^-$ $\sqrt{s}$=91 GeV)=1.84 GeV/$fm^3$.
\vspace{0.3cm}

\noindent
 The systematic error (which in this case is dominated by the
 uncertainty in the $R_T$ determination) is about 50\%.
\vspace{0.3cm}
\vspace{0.3cm}

\noindent
\underline{\bf 2. $p + \overline{p}$ at $\sqrt{s}$=1.8 TeV} \hspace{0.2cm}
For $p + \overline{p}$ we use the same method, while we use as
transverse radius the radius of the nucleon of 0.8 fm.
We use again pions, kaons, nucleons and antinucleons,
and the mean transverse momenta measured in \cite{Gutay}.
We derive the $dE_T/dy(ycm$  from the measured charged
total multiplicity per unit rapidity at midrapidity,
using the ratios and mean transverse momenta 
measured as a function of charged multiplicity in \cite{Gutay}.
No extrapolation factor to midrapidity is needed here.
We find 

\begin{tabular}{lll}
$\epsilon_{\ in} \ (dN_c/d\eta=19)$ & = & 7.61 GeV/$fm^3$,
\\
$\epsilon_{\ in} \ (dN_c/d\eta=15)$ & = & 5.82 GeV/$fm^3$,
\\
$\epsilon_{\ in} \ (dN_c/d\eta=11.5)$ & = & 4.31 GeV/$fm^3$,
\\
$\epsilon_{\ in} \ (dN_c/d\eta=5.4)$ & = & 2.03 GeV/$fm^3$,
\\
$\epsilon_{\ in} \ (dN_c/d\eta=3)$ & = & 0.772 GeV/$fm^3$.
\end{tabular}

\noindent
In reference \cite{Gutay} the $\epsilon_{\ in}$ for $dN_c/d\eta=15$ 
of 3 GeV/fm$^3$ is calculated 
using  pions only, and therefore  is lower than our estimate\footnote{We
find the same $\epsilon_{\ in}$ as reference \cite{Gutay}, when
we use only pions.}.
\vspace{0.3cm}

\noindent
\underline{\bf 3. $p+ \overline{p}$ $\sqrt{s}$=900 GeV} \hspace{0.2cm}
For $p+ \overline{p}$ at $\sqrt{s}$=900 GeV 
we use the multiplicities from \cite{becattini} and the mean transverse momenta
for $p+ \overline{p}$ at $\sqrt{s}$=1.8 TeV from \cite{Gutay}.
We took the mean multiplicity per unit rapidity without extrapolation
factor A to midrapidity since we don't know the shape of the rapidity
distributions.
We find
\vspace{0.3cm}

$\epsilon_{\ in}$($p+ \overline{p}$ $\sqrt{s}$=900 GeV)= 1.23 GeV/$fm^3$
\vspace{0.3cm}

\noindent
The pseudorapidity distribution of charged particles produced in
$p + \overline{p}$ collisions at $\sqrt{s}$=1.8 TeV is relatively
flat over  6 rapidity units as measured by CDF \cite{atlas}.
However this may differ at the lower $\sqrt{s}$ of 900 GeV.
We use the extrapolation factor A=1.49 (=$dN/dy(ycm)/dN/dy(mean)$) from
the measured $dN/dy$ distribution
in \cite{tasso}.
The resulting $\epsilon_{\ in}$ has an uncertainty of about 50\%.
\vspace{0.3cm}

\noindent
\underline{\bf 4. Au+Au $\sqrt{s}$=130 GeV} \hspace{0.2cm}
For  Au+Au at $\sqrt{s}$=130 GeV we used the mean $p_T$ from
$p+ \overline{p}$  collisions at $\sqrt{s}$=1.8 TeV and the
particle multiplicities at midrapidity
(negative hadrons, total charged multiplicity, nucleons estimated 
from $p - \overline{p}$ $\sim$ $N_{charged} - N_{negatives}$,
as well as antinucleons using the $\overline{p}/p$ ratio)
from \cite{hepex0007036,star}. 
\vspace{0.3cm}

\noindent
With the above method ($\alpha$) (equations \ref{bjorken} , \ref{et_tot}) 
we calculated the $\epsilon_{\ in}$ for several reactions.
We summarize the results here:
\vspace{0.3cm}

$\epsilon_{\ in}$ (Au+Au $\sqrt{s}$=4.9 GeV) =  0.86 GeV/$fm^3$,
with $m_T-m_0$ instead of $m_T$ for nucleons
\vspace{0.3cm}

$\epsilon_{\ in}$(Si+Au $\sqrt{s}$=5.4 GeV) = 0.53  GeV/$fm^3$, 
with $m_T-m_0$ instead of $m_T$ for nucleons
\vspace{0.3cm}

In the following we used $m_T$ for nucleons too
\vspace{0.3cm}

$\epsilon_{\ in}$(Au+Au $\sqrt{s}$=4.9 GeV) =  1.70 GeV/$fm^3$,
\vspace{0.3cm}

$\epsilon_{\ in}$(Si+Au $\sqrt{s}$=5.4 GeV) = 1.53  GeV/$fm^3$, 
\vspace{0.3cm}

$\epsilon_{\ in}$(S+S $\sqrt{s}$=19 GeV) =  1.21 GeV/$fm^3$, 
\vspace{0.3cm}

$\epsilon_{\ in}$(Pb+Pb $\sqrt{s}$=17 GeV) =  2.4 GeV/$fm^3$, 
\vspace{0.3cm}

$\epsilon_{\ in}$(Au+Au $\sqrt{s}$=130 GeV) =  6.34 GeV/$fm^3$, 
\vspace{0.3cm}

$\epsilon_{\ in}$($p+\overline{p}$ $\sqrt{s}$=900 GeV) =  1.23 GeV/$fm^3$, 
\vspace{0.3cm}

$\epsilon_{\ in}$($p+\overline{p}$ $\sqrt{s}$=1.8 TeV most central) =  
7.61 GeV/$fm^3$,
\vspace{0.3cm}

$\epsilon_{\ in}$($p+\overline{p}$ $\sqrt{s}$=1.8 TeV most peripheral) =  
0.77 GeV/$fm^3$, 
\vspace{0.3cm}

$\epsilon_{\ in}$(p+p $\sqrt{s}$=17 GeV) =  0.42 GeV/$fm^3$, 
\vspace{0.3cm}

$\epsilon_{\ in}$($e^+ + e^-$  $\sqrt{s}$=91 GeV) =  1.84 GeV/$fm^3$, 
\vspace{0.3cm}
\vspace{0.3cm}

\noindent
The initial energy density has been estimated by experimenters 
for some reactions shown here,
using an other method ($\beta$), namely
the Bjorken formula (equation \ref{bjorken})
and the measured transverse energy :

\begin{equation}
E_T({lab}) \ = \ (E sin \theta)_{lab}
\label{esin}
\end{equation}

\noindent
(method ($\beta$))
\vspace{0.3cm}

\noindent
where E is the total energy measured with e.g. calorimeters and $\theta$ is
the angle to the incident beam direction.
The resulting values, (used in figures 1,  2 and 5 in the following section),
are:
\vspace{0.3cm}

$\epsilon_{\ in}$(Pb+Pb $\sqrt{s}$=17 GeV) =  3.2 GeV/$fm^3$ \cite{na49_et}\footnote{The first three values here are valid for head-on collisions \cite{na49_et}.},
\vspace{0.3cm}

$\epsilon_{\ in}$(S+Au) $\sqrt{s}$=19 GeV) =  2.6 GeV/$fm^3$ 
\cite{na49_et,na35_et},
\vspace{0.3cm}

$\epsilon_{\ in}$(S+S $\sqrt{s}$=19 GeV) =  1.3 GeV/$fm^3$ 
\cite{na49_et,na35_et},
\vspace{0.3cm}

$\epsilon_{\ in}$(Au+Au $\sqrt{s}$=4.9 GeV) =  1.3 GeV/$fm^3$ 
\cite{nuclex9803015},
\vspace{0.3cm}

$\epsilon_{\ in}$(Si+Au $\sqrt{s}$=5.4 GeV) = 0.9  GeV/$fm^3$  
(see \cite{Sonja} and references therein).
\vspace{0.3cm}

\noindent
We estimate the maximal initial energy density with a third method
($\gamma$), taking
the nuclear energy density of two overlapping nuclei 
2 $\epsilon_A$ times the $\gamma$ factor of the colliding particles
in the center of mass minus one:

\begin{equation}
\epsilon_{\gamma} \ = \ 2 \ \epsilon_A \ (\gamma -1)  
\label{gamma}
\end{equation}

\noindent
(method ($\gamma$))
\vspace{0.3cm}

\noindent
with $\gamma \ = \ (\sqrt{s}/2)/m_{nucleon}$, and
$\epsilon_{A, small}$=0.179 GeV/fm$^3$ for small nuclei and
$\epsilon_{A, big}$=0.138 GeV/fm$^3$ for large nuclei 
is the normal nuclear matter density.
The value in equation \ref{gamma}  multiplied by the stopping power
gives an estimate of the initial energy density.
\vspace{0.3cm}

\noindent
We apply this method to the Si+Au and Au+Au collisions at 
$\sqrt{s} \sim$ 5 GeV, Ni+Ni collisions at $\sqrt{s}$=2.6 GeV,
and Pb+Pb collisions at $\sqrt{s}$=17 GeV, to calculate the maximal achieved
initial energy density, yielding:
\vspace{0.3cm}

$\epsilon_{\ in}$(Au+Au $\sqrt{s}$=4.9 GeV) =  0.44 GeV/$fm^3$,
\vspace{0.3cm}

$\epsilon_{\ in}$(Si+Au $\sqrt{s}$=5.4 GeV) =  0.67 GeV/$fm^3$, 
\vspace{0.3cm}

$\epsilon_{\ in}$(Ni+Ni $\sqrt{s}$=2.6 GeV) =  0.276 GeV/$fm^3$, 
\vspace{0.3cm}

$\epsilon_{\ in}$(Pb+Pb $\sqrt{s}$=17 GeV) = 2.25  GeV/$fm^3$.
\vspace{0.3cm}

\vspace{0.3cm}
\vspace{0.3cm}

Few comments:

\begin{enumerate}

\item
A problem with method ($\alpha$) is,
that in general it underestimates slightly the $E_t$, since not
all particles are taken into account.

\item
A general problem is with equation \ref{esin} of method ($\beta$):
For particles produced near midrapidity $E \times sin{ \theta}$ is
approximately
$p_T$ not $E_T$, for $\gamma_{cm} >> 1$,
whereas $E \times sin{ \theta}$ is approximately $m_T$ for particles
moving nonrelativistically in the lab frame.

\item
A problem with method ($\gamma$) is, that  it assumes geometrical and not
dynamical compression, possibly slightly underestimating the $\epsilon_{\ in}$.

\item
The methods ($\alpha$) and $\beta$)
should in principle give similar results.
Our $\epsilon_{\ in}$(method ($\alpha$)) 
agrees well with the calorimetric estimation
of NA35 for  S+S at $\sqrt{s}$=19 GeV 
and of NA49 for Pb+Pb at $\sqrt{s}$=17 GeV \cite{na35_et,na49_et},
for the 5\% $\sigma_{tot}$ 
centrality trigger.
We compare to $\epsilon_{\ in}$(Pb+Pb) $\sim$ 0.77 $\times$ 3.2 GeV/$fm^3$=
2.46 GeV/$fm^3$,
with 
0.77 = $(dE_T/d\eta) (5\% \sigma_{tot}) / (dE_T/d\eta)$ (head on collisions)
from \cite{na49_et}.
This correction is needed because the $\epsilon_{\ in}$=3.2 GeV/$fm^3$
is estimated for head on collisions, while our calculation uses $m_T$ and
d/dy's taken with the 5\% $\sigma_{tot}$ centrality trigger.
\\
Additionally, also method ($\gamma$), 
- and therefore all three methods ($\alpha$), ($\beta$) and ($\gamma$)-,
give   
similar results for Pb+Pb collisions at $\sqrt{s}$=17 GeV.
\vspace{0.3cm}

\item
We find however different $\epsilon_{\ in}$ values for the low $\sqrt{s}$ 
reactions Si+Au and Au+Au at 5.4 and 4.9 GeV,
for each one of the above methods with a scatter of a factor of two
or more.
Also 
the maximal estimated $\epsilon_{\ in}$ with method ($\gamma$)
is lower than the
results from methods ($\alpha$) and ($\beta$).
We therefore conclude that the Bjorken estimate may not be adequate for
low  energy reactions.
This estimate was not meant to be used in the nonrelativistic
regime.
The method ($\gamma$) seems more adequate for these systems.
Nevertheless we show two distinct cases in the following figures.

\noindent
Using method ($\gamma$) the $\epsilon_{\ in}$ is slightly higher for
Si+Au at 5.4 GeV than for Au+Au at 4.9 GeV  $\sqrt{s}$,
unlike the results  of methods ($\alpha$) and ($\beta$).
The stopping power which is missing in the
equation \ref{gamma}, can however hardly be very different
for these two reactions, but may have an influence on
the ordering of the $\epsilon_{\ in}$ in the above two reactions.

\end{enumerate}
\vspace{0.3cm}

\noindent
{\bf Systematic error estimation summary}
Comparing results from methods ($\alpha$) and ($\gamma$) we
estimate the systematic error in Si+Au and Au+Au collisions at
$\sqrt{s}$=5.4 and 4.9 GeV  to be about 50\%, mainly resulting
from the difficulty in applying the Bjorken formula.
For the higher energy nucleus+nucleus collisions the systematic error is smaller 
$\sim$ 30\% \cite{hepph0004138} resulting from the estimation of
$E_T$.
The systematic error for $e^+ + e^-$ is about 50\%, resulting
from uncertainties in the transverse area definition.
The systematic error for
$p+p$ colllisions at $\sqrt{s}$=900 GeV results from 
the $E_T$ definition and the extrapolation to midrapidity
and is estimated to be about 50\% .
For  $p + \overline{p}$ collisions at the Tevatron, the uncertainty
comes mainly from not taking all particles into account with method ($\alpha$)
and is  about 30\%.
The systematic error on the $\epsilon_{\ in}$ estimate for Ni+Ni at
$\sqrt{s}$=2.6 GeV
is  50\%, resulting from comparison of our $\epsilon_{\ in}$ value
to model calculations \cite{rbuu}.

\vspace{0.5cm}
\vspace{0.5cm}

\subsection{The combined results: T, $\lambda_s$ at zero fugacities}

\begin{figure}[ht]
\begin{center}
\mbox{\epsfig{file=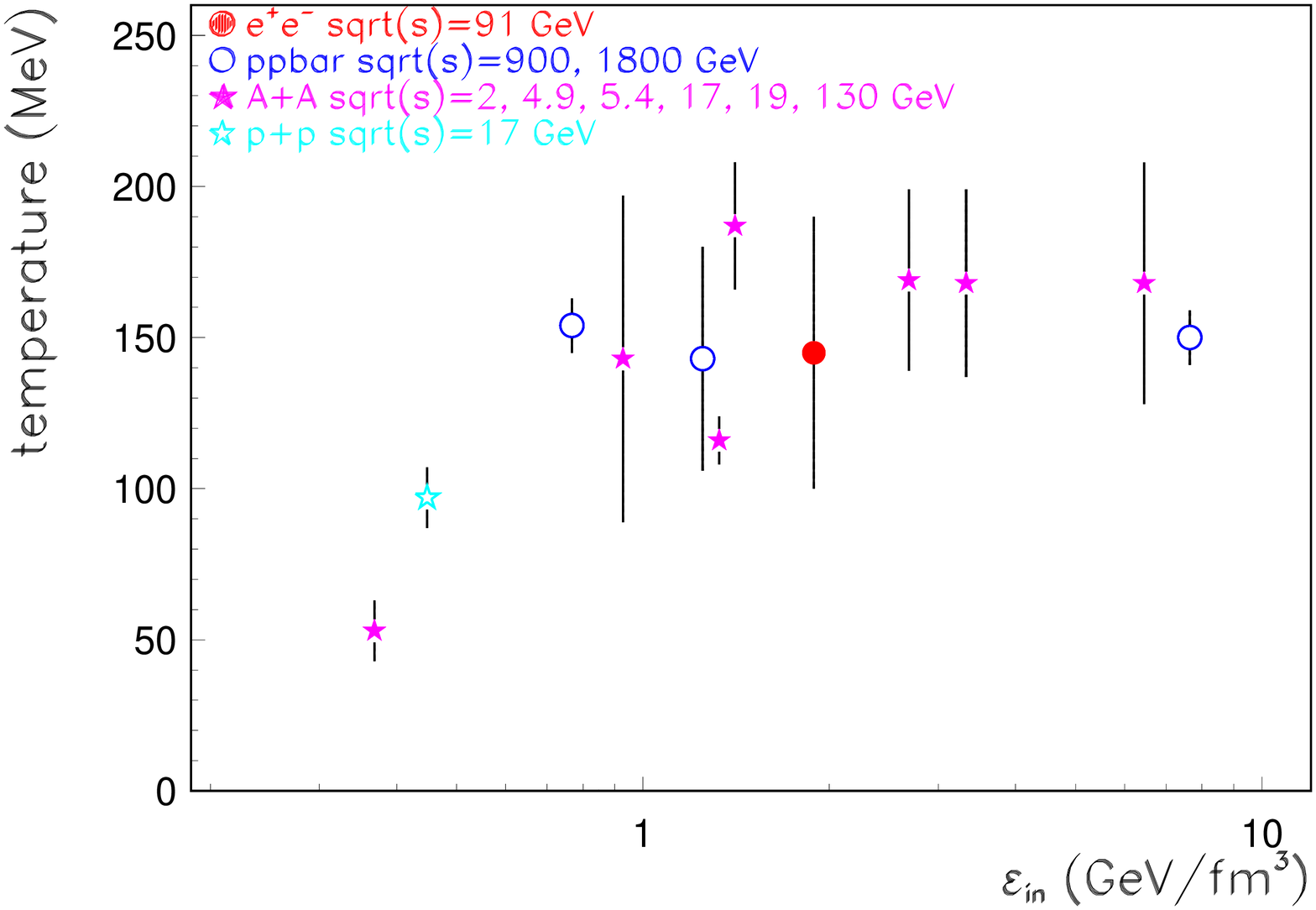,width=120mm}}
\end{center}
\caption{
Temperature at  chemical freeze-out extrapolated to zero fugacities along
an isentropic path, as a function of
the initial energy density for several nucleus+nucleus, hadron+hadron and
lepton+lepton collisions. 
We demand for the fits confidence level $>$ 1\%.
}
\label{t}
\end{figure}

\begin{figure}[ht]
\begin{center}
\mbox{\epsfig{file=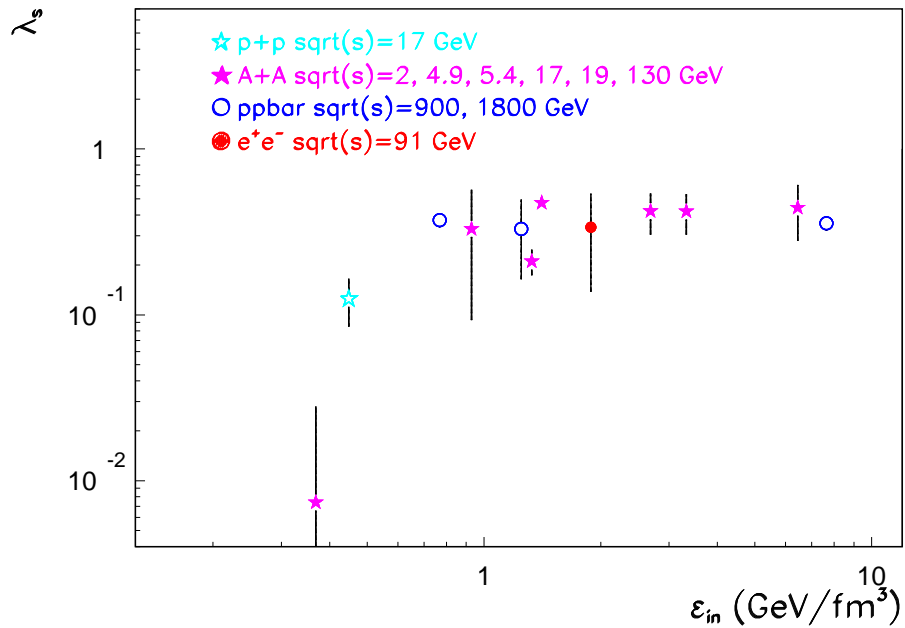,width=120mm}}
\end{center}
\caption{
The number of strange antiquarks over the mean number of light ones
at  chemical freeze-out and for zero fugacities 
as a function of
the initial energy density for several nucleus+nucleus, hadron+hadron and
lepton+lepton collisions.
We demand for the fits confidence level $>$ 1\%.
}
\label{l}
\end{figure}

Figures 
\ref{t} and \ref{l} show the resulting temperature and the $\lambda_s$ 
factor extrapolated to zero fugacities along isentropic paths,
as a function of the initial energy density.
The latter is taken from the experimentally measured 
transverse energy, when available, that is, for 
Si+Au $\sqrt{s}$=5.4 GeV, Au+Au $\sqrt{s}$=4.9 GeV,
S+S $\sqrt{s}$=19 GeV, S+A $\sqrt{s}$=19 GeV and
Pb+Pb $\sqrt{s}$=17 GeV.
For the  remaining colliding  systems, where no initial energy density
estimation is available, we use our method ($\alpha$) based on equation 
\ref{et_tot}, exept for the Ni+Ni system where we use the
method ($\gamma$).
The fact that the initial energy density estimation in figures \ref{t} and
\ref{l}, was not done
with the same method, introduces an additional systematic
error on the $\epsilon_{\ in}$ scale.
To reduce this uncertainty and show the systematic error of
the initial energy density estimation we use in the following
the $\epsilon_{\ in}$  estimated for 1.) high $\sqrt{s}$ ($>$10 GeV)
by equations \ref{bjorken} and \ref{et_tot}, and for 2.) low $\sqrt{s}$ ($<$10
GeV) by equation \ref{gamma}.
\vspace*{0.3cm}

\begin{figure}[ht]
\begin{center}
\mbox{\epsfig{file=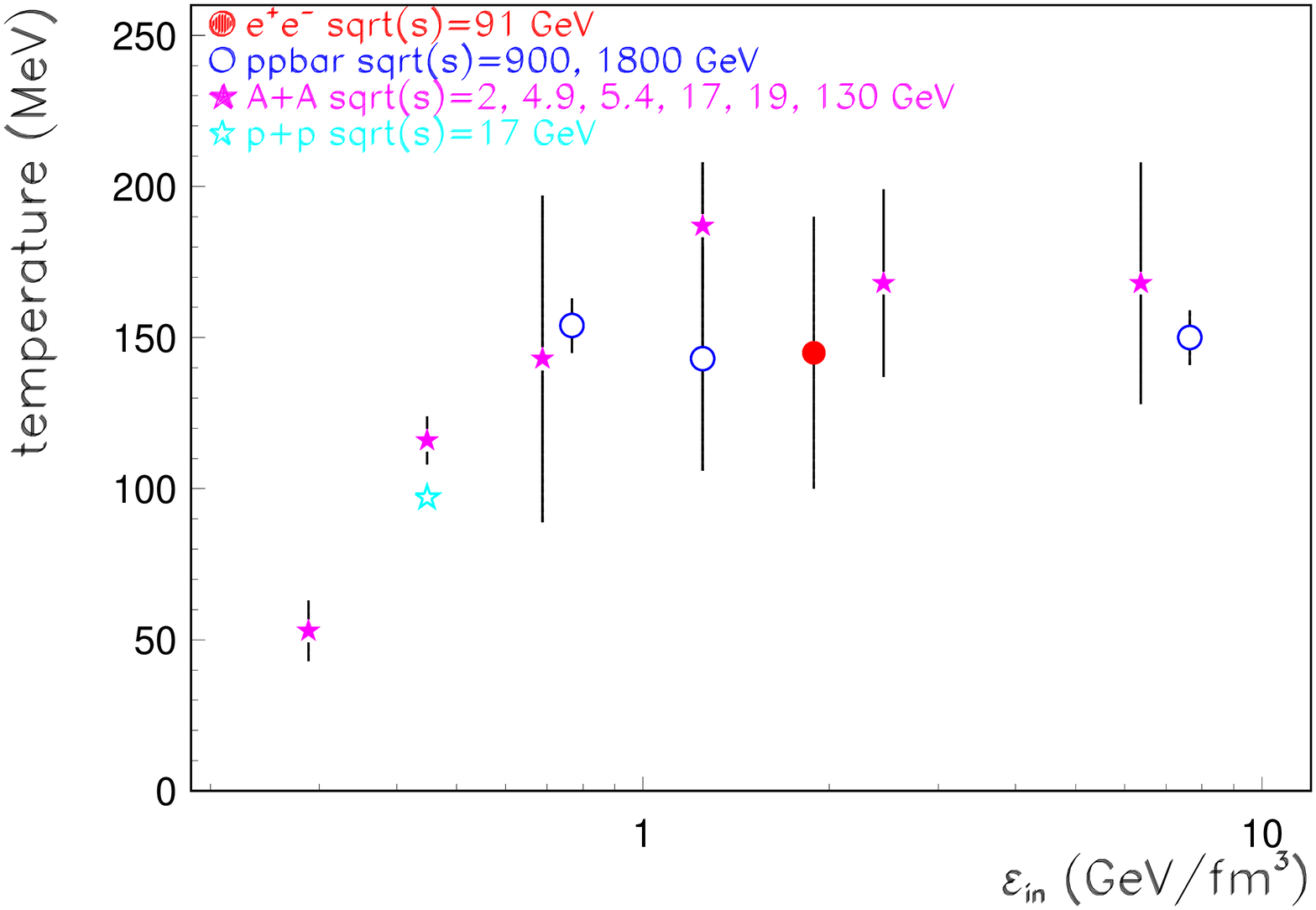,width=120mm}}
\end{center}
\caption{
Temperature at  chemical freeze-out and for zero fugacities as a function of
the initial energy density for several nucleus+nucleus, hadron+hadron and
lepton+lepton collisions. 
The initial energy density has been estimated
in a different way as the one in figures \ref{t} and \ref{l} (see text).
We demand for the fits confidence level $>$ 1\%.
}
\label{t2}
\end{figure}

\begin{figure}[ht]
\begin{center}
\mbox{\epsfig{file=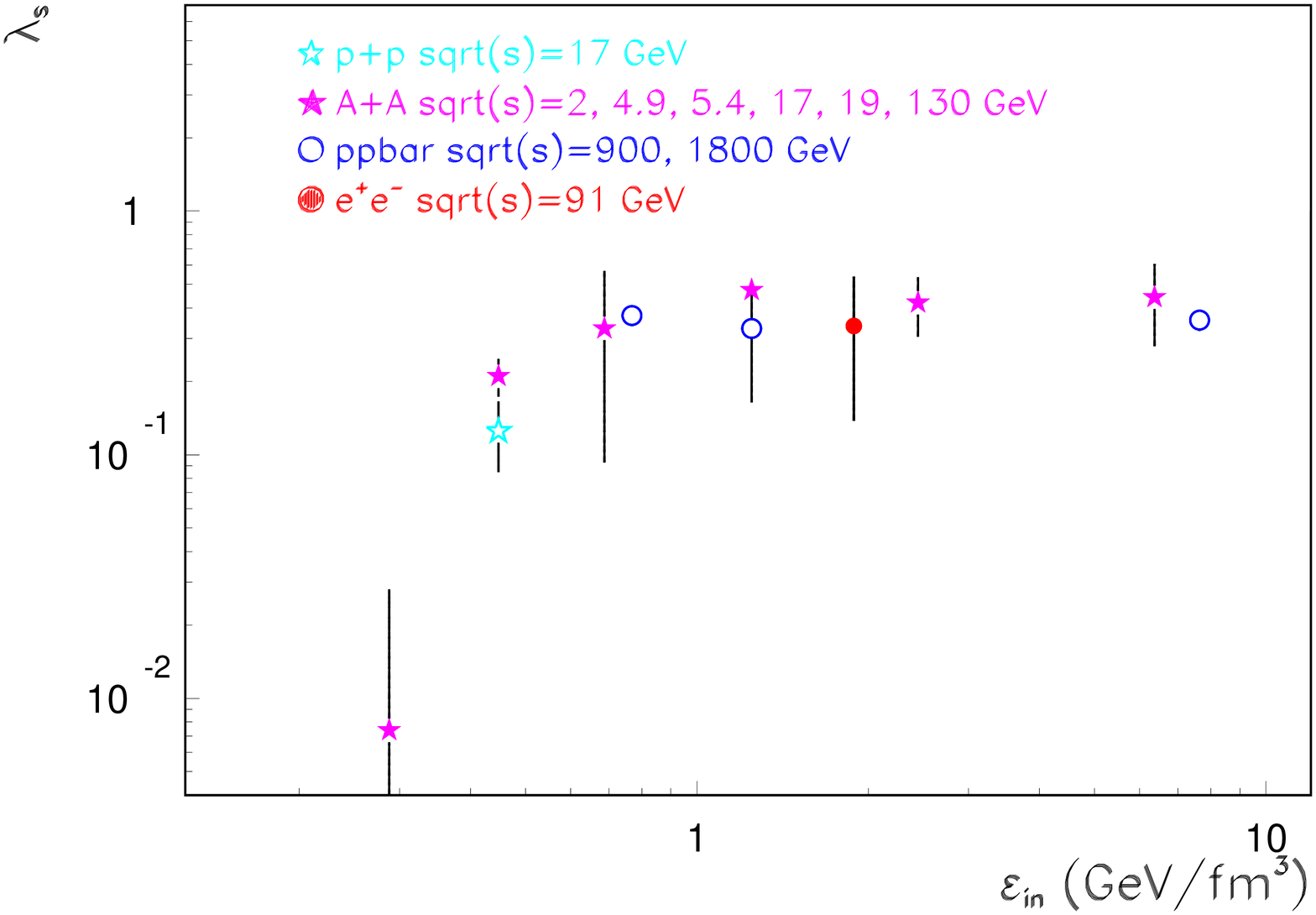,width=120mm}}
\end{center}
\caption{
The number of strange antiquarks over the mean number of light ones
at  chemical freeze-out and for zero fugacities 
as a function of
the initial energy density for several nucleus+nucleus, hadron+hadron and
lepton+lepton collisions.
The initial energy density has been estimated
in a different way as the one in figures \ref{t} and \ref{l} (see text).
We demand for the fits confidence level $>$ 1\%.
}
\label{l2}
\end{figure}

\noindent
Figures 
\ref{t2} and \ref{l2} show therefore the temperature and the $\lambda_s$ 
factor again, extrapolated to zero fugacities along isentropic paths,
as a function of the initial energy density, with a different
estimation of the latter.
In particular, the initial energy density
is not taken from the experimentally measured 
transverse energy,
but is estimated using our method ($\alpha$) based on equation
\ref{et_tot}, with the exeption of the data at
low $\sqrt{s}$ ($<10$ GeV), that is,  for Ni+Ni at
$\sqrt{s}$=2.6 GeV, Si+Au at $\sqrt{s}$=5.4 GeV and Au+Au at
$\sqrt{s}$=4.9 GeV, where we used equation \ref{gamma}, which
as discussed in the previous section, seems more adequate for
the low energies as the Bjorken estimate.
\vspace*{0.3cm}

\noindent
Figure \ref{t_k} (a) shows the number density of kaons
as a function of the initial energy density from reference
\cite{Sonja}, while in figure \ref{t_k} (b) the
 temperature extrapolated to zero fugacities along isentropic paths,
is shown as a function of the initial energy density.
In figure \ref{t_k} (a), the collision systems 
p+p  at $\sqrt{s}$=17 GeV, S+S at $\sqrt{s}$=19 GeV,
Pb+Pb at $\sqrt{s}$=17 GeV, Au+Au at $\sqrt{s}$=4.9 GeV and
Si+Au at $\sqrt{s}$=5.4 GeV are shown.
In figure \ref{t_k} (b) all the above reactions are shown
and additionally the remaining analysed heavy ion collisions
in the last section, namely S+A at $\sqrt{s}$=19 GeV,
Au+Au at $\sqrt{s}$=130 GeV and Ni+Ni at  $\sqrt{s}$=2.6 GeV.
The initial energy density of the common reactions
displayed in figures \ref{t_k} (a) and (b), -that is, of all
reactions shown in \ref{t_k} (a)-, 
is defined in the same way, to allow their direct comparison.
In particular, the initial energy densities for
all heavy ion systems is taken from the experimental
calorimetric measurements
(method $\beta$ above) when these are available.
For Au+Au at $\sqrt{s}$=130 GeV we use our estimate with method ($\alpha$).
For p+p  at $\sqrt{s}$=17 GeV we used the estimate from reference
\cite{Sonja} of 
\vspace*{0.3cm}

$\epsilon_{\ in}$(p+p  at $\sqrt{s}$=17 GeV) from \cite{Sonja}=0.85 GeV/fm$^3$
\vspace*{0.3cm}

\noindent
This is found in \cite{Sonja}
by a) estimating the dependence of the $\epsilon_{\ in}$ for
Pb+Pb collisions at $\sqrt{s}$=17 GeV on the number of participant nucleons N
and b) extrapolating this function to N=2.
\vspace*{0.3cm}

\noindent
The rise and subsequent saturation seen in the kaon number density
below and above $\epsilon$=1.3 GeV/fm$^3$ (figure \ref{t_k} (a)),
shows a clear relation to
 the same behaviour seen in the
temperature (figure \ref{t_k} (b)) as a function of the 
initial energy density.
\vspace*{0.3cm}

\begin{figure}[ht]
\begin{center}
\mbox{\epsfig{file=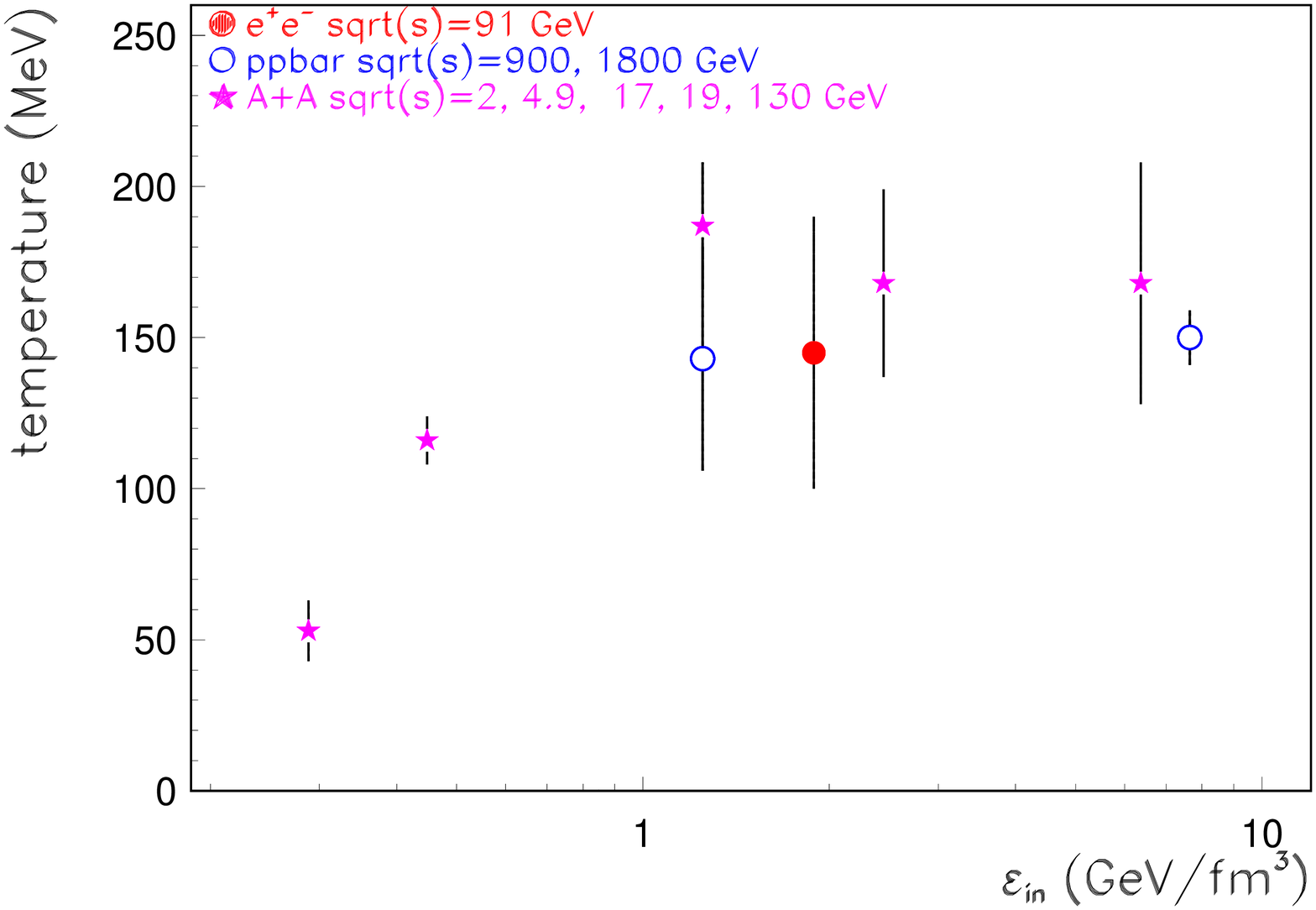,width=120mm}}
\end{center}
\caption{
Temperature at  chemical freeze-out and for zero fugacities as a function of
the initial energy density for several nucleus+nucleus, hadron+hadron and
lepton+lepton collisions. 
The initial energy density has been estimated
in a different way as the one in figures \ref{t} and \ref{l} (see text).
We demand for the fits confidence level $>$ 10\%.
}
\label{t3}
\end{figure}

\begin{figure}[ht]
\begin{center}
\mbox{\epsfig{file=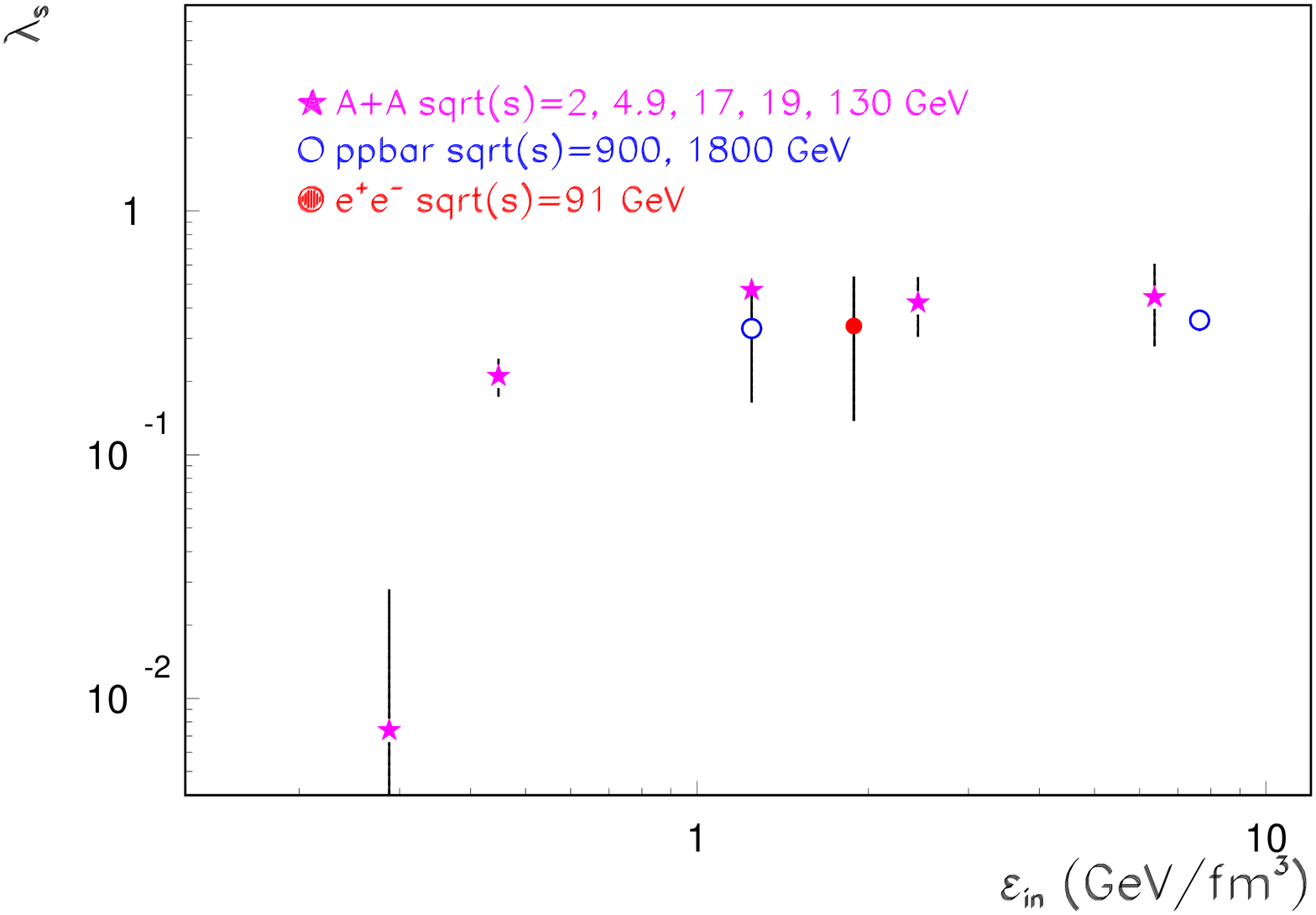,width=120mm}}
\end{center}
\caption{
The number of strange antiquarks over the mean number of light ones
at  chemical freeze-out and for zero fugacities 
as a function of
the initial energy density for several nucleus+nucleus, hadron+hadron and
lepton+lepton collisions.
The initial energy density has been estimated
in a different way as the one in figures \ref{t} and \ref{l} (see text).
We demand for the fits confidence level $>$ 10\%.
}
\label{l3}
\end{figure}

\begin{figure}[ht]
\begin{center}
\mbox{\epsfig{file=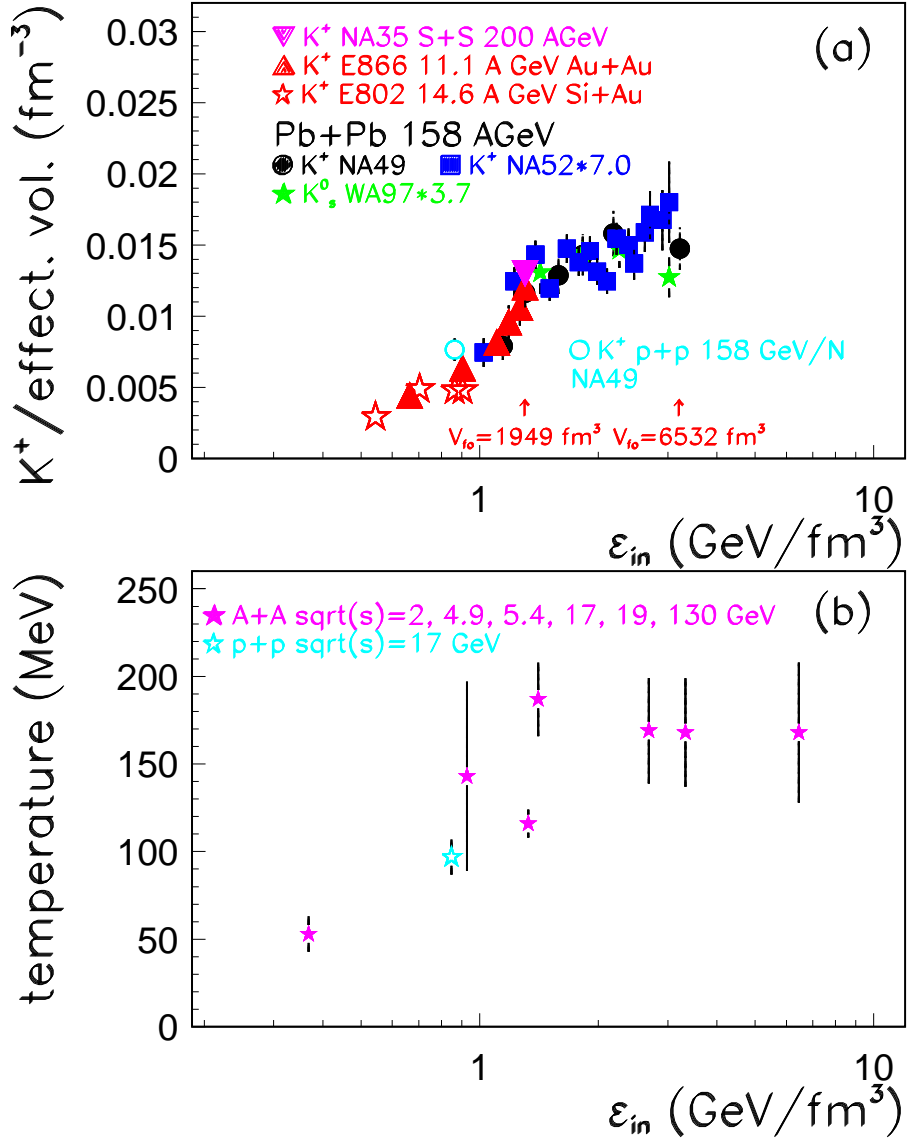,width=120mm}}
\end{center}
\caption{
(a) Kaon yield per interaction over the source volume at the thermal
freeze-out as a function of the initial energy density 
($\epsilon_{\ in}$) from reference \protect\cite{hepph0004138,Sonja}.
(b) 
Temperature at  chemical freeze-out and for zero fugacities as a function of
the initial energy density for several nucleus+nucleus, hadron+hadron and
lepton+lepton collisions. 
All collision systems shown in figure (a), are displayed in
figure (b) using the same initial energy density as defined in
\protect\cite{hepph0004138,Sonja}.
We demand for the fits confidence level $>$ 1\%.
}
\label{t_k}
\end{figure}

\noindent
It is apparent from figures \ref{t}, \ref{l}, \ref{t2}, 
\ref{l2}, \ref{t_k} and the present discussion,
that  a precise calculation of the 
critical energy density affecting all figures (\ref{t}-\ref{t_k}), needs
more experimental data in the $\epsilon$ region around 
1 GeV/fm$^3$. 
The errors on the temperature (figures \ref{t}, \ref{t2})
are large around
this region.
An other uncertainty arises from the determination of the
initial energy density especially at low $\sqrt{s}$ where the Bjorken
estimate may not be adequate.
Our best estimate of the initial energy density, is shown
in figures \ref{t2}, \ref{l2}, \ref{t3}, \ref{l3}.
In figures \ref{t}, \ref{l}, \ref{t2},  \ref{l2},\ref{t_k}
we show results demanding a confidence level greater than 1\%.
In figures \ref{t3} and \ref{l3} we show the results demanding
a confidence level greater than 10\%.

\clearpage
\newpage

\section{Conclusions}

\noindent
We have discussed in section 2 the dependence of the critical temperature
in the QCD phase transition on the vacuum pressure including in addition
modes of noninteracting hadron resonances for u,d,s flavors
of light quarks as defining the hadronic phase.
Equating the pressure of this hadronic phase to the pressure of the plasma
phase, represented by noninteracting u,d,s  quarks and antiquarks and gluons
as shown in figure 1, we obtain for zero fugacities
$T_{\ cr} \ = \ 194 \ \pm \ 18 \ MeV$ .

\noindent
We have performed a thermal analysis of yields in multiparticle production
for 13 reactions summarized in table \ref{chisquare} and discussed
in detail in section 3. The intensive
thermal parameters ( temperature and baryon as well as strangeness fugacity )
together with an error estimate are used to extrapolate the 
state associated with the chemical freeze-out of each reaction, studied
along curves of equal entropy- , energy- and number-density,
to zero fugacities.

\noindent
We  represent the so obtained states by two intensive parameters 

$ T $ and $\lambda_{\ s} \ = \ \frac{2 \left \langle \overline{s} \right \rangle}
{ \left \langle \overline{u} \right \rangle \ + \ \left \langle \overline{d} 
\right \rangle}$ ,

\noindent
i.e. temperature and the ratio of antistrange to nonstrange (valence)
antiquark abundances. These two quantities as functions of the initial
hadronic energy density achieved in each reactions are displayed
in figures \ref{t} - \ref{l2}. The initial energy densities are
estimated in subsection 3.4. The resulting error (horizontal error in
figures \ref{t} - \ref{t_k}) is of the order of 50\% around
$\epsilon_{crit}$, for the $e^+ + e^-$ collisions at $\sqrt{s}$=91 GeV
and for $p + \overline{p}$ collisions at $\sqrt{s}$=900 GeV,
and approximately 30\% at higher $\epsilon_{\ in}$.
\vspace{0.3cm}

\noindent
The reactions with an estimated
confidence level above 10\% were retained, which
fall within the errors into two groups :

\begin{description}
\item 
group I with $\varepsilon_{\ in} \ \stackrel{>}{\sim} \ 1 GeV/fm^{\ 3}$

\begin{enumerate}
\item 
central Au+Au collisions at RHIC $\sqrt{s} \ = 130 \ A \ GeV$

\item 
central Pb+Pb collisions at  $\sqrt{s} \ = 17 \ A \ GeV$

\item
central S+A (A=Pb,Au,W) collisions $\sqrt{s} \ = 19 \ A \ GeV$

\item
central S+S collisions at $\sqrt{s} \ = 19 \ A \ GeV$

\item 
$e^+$+ $e^-$ collisions at LEP  $\sqrt{s} \ = 91.19 \ GeV$

\item
$p+\overline{p}$ collisions at $\sqrt{s} \ = \ 900 \ GeV$

\item
central $p+\overline{p}$ collisions at $\sqrt{s} \ = \ 1.8 \ TeV$ 

\end{enumerate}

\item 
group II with $\varepsilon_{\ in} \ \stackrel{<}{\sim} \ 1 GeV/fm^{\ 3}$

\begin{enumerate}
\setcounter{enumi}{7}

\item 
central Au+Au collisions at $\sqrt{s} \ = 4.9 \ GeV$  

\item 
Ni+Ni collisions at GSI $\sqrt{s} \ = 2.8 \ A \ GeV$ 
\end{enumerate}
\end{description}

\noindent
Both quantities T and $\lambda_{s}$ saturate to constant values

\begin{equation}
\label{satur}
\begin{array}{l} 
T_{\ lim} \ = 155 \pm 6 \pm 20 \ MeV \ (syst)
\hspace*{0.2cm} ; \hspace*{0.2cm} 
\lambda_{\ lim} \ =  0.365 \pm 0.033 \pm 0.07 \ (syst)
\end{array}
\end{equation}

\noindent
above the dividing energy density $\varepsilon_{\ in} \ \sim \ 1 \
GeV/fm^{\ 3}$.
Within the errors this is compatible with the critical energy density
$\varepsilon_{\ crit} \ \sim \ 1 GeV/fm^{\ 3}$ obtained in lattice
QCD \cite{lattice} as well as with the results on the critical parameters
obtained in section 2 ( eq. \ref{eq:12} ).
This latter agreement and the extension of the work in
\cite{PM} to include excited hadronic degrees
of freedom is a new result.
\vspace{0.3cm}

\noindent
This limiting temperature is expected to be somewhat below the critical one,
because hadronization is not an instantaneous process. The mean
values show a difference in temperature of 40 MeV but the errors
mainly due to the approximations of free quasiexcitations and experimental
errors are large adding linearly also amount to $\sim$ 40 MeV. 
\vspace{0.3cm}

\noindent
We estimate the systematic error on the limiting $T$, $\lambda_s$
values, fitting the high $\epsilon_{\ in}$ region of the 
 figures \ref{t2}, \ref{l2} using once a line fit and once
a horizontal line fit, varying in both cases the 
number of bins taken, and estimating the deviations
of the mean values resulting from each of the two fits.
The error is $\sim$ 20 MeV for the temperature and 0.07 for the
$\lambda_s$.
\vspace{0.3cm}

\noindent
We interpret this saturation as characteristic for all reactions which
before chemical freeze-out have passed through the quark gluon plasma
phase, as opposed to those which remained throughout in the hadronic phase.
\vspace{0.3cm}

\noindent
We note that due to the large errors the dividing line as drawn at
$\varepsilon_{\ in} \ \sim \ 1 GeV/fm^{\ 3}$ is subject to a corresponding
uncertainty.

\noindent
The saturation phenomenon observed here is in our interpretation
confirmed by the same phenomenon derived for  the kaon number density
as a function of the initial energy density by one of us 
\cite{hepph0004138,Sonja} (figure
\ref{t_k}).

\noindent
The agreement of the theoretical thermodynamic model with experimental
ratios of identified particle multiplicities is quantified in
the errors as discussed in section 3. We emphasize that agreement
cannot be perfect due to several obvious and less obvious facts :

\begin{description}
\item
a ) The rapidity and transverse momentum distributions do not agree with
 the spherically symmetric distributions in the thermodynamic description.

\item
b ) There may well exist kinematically distinct thermal systems, as is indicated
by the phenomenon of longitudinal and transverse flow \cite{flow}. 

\item
c) The target and projectile diffractive regions may form distinct
   further thermal subsystems characterized by a different
   temperature than the particles produced near midrapidity \cite{Sonja2}.
\end{description}

\noindent
The results derived here do indicate in our interpretation a high degree
of thermalization of the systems studied. This implies the clear indication,
that the quark gluon plasma phase is part of the hadronization process
characteristic for hadronic, leptonic and nuclear reactions above
the critical energy density.

\noindent
The universal excitation of the quark gluon plasma phase in
nuclear {\it and} hadronic as well as leptonic multiparticle production is a new result.
\vspace{0.3cm}

\noindent
The behaviour of strangeness production as reflected
by the parameter $\lambda_{\ s}$ also reveals a new aspect:
the saturation phenomenon corresponds to a strangeness enhancement relative
to states with low initial energy density.
No 'strangeness enhancement' is seen for heavy ion collisions relative to 
 hadronic and leptonic reactions, when compared at the same initial
energy density and zero fugacities.
This enhancement  arises only if systems
with very different thermal properties in particular different
fugacities are compared to one another at the same $\sqrt{s}$, therefore
 at different energy density when  nucleus nucleus collisions are
 compared to elementary particle collisions.
 \vspace*{0.3cm}

\noindent
LHC results for central p+p and Au+Au collisions at $\sqrt{s}$= 14, 5.5 TeV
will be important for the confirmation of the above picture.
\\
\noindent
We conclude by proposing further experimentation in the following areas :

\begin{description}
\item
i ) study of p+ $\overline{p}$ collisions at the Tevatron, with centrality
selection on transverse energy, similar to nuclear collisions, and
possibly with an improved particle identification option including
the fragmentation regions. 

\item
ii ) in conjunction with i) an extension to compare open and closed c+ $\overline{c}$
production.

\item
iii) within the program of heavy ion collisions at RHIC and at the SPS,
allready under way, a dedicated study 
in the neighborhood of 
$\epsilon_{in, crit}$ = 1 GeV/fm$^3$,
e.g. for $\epsilon_{in}$=(0.6-1.5) GeV/fm$^3$, in nucleus+nucleus,
p+A and p+p collisions, using centrality selection.
\end{description}

\subsection*{Acknowledgements}

We thank K. Pretzl, W. Ochs, F. Niedermeyer and H. Balsiger for stimulating and 
fruitfull discussions and critical comments. We thank W. Ochs for
very helpful informations concerning jet physics.
We thank the members of the theory division of CERN, where part of this work 
was done, for their hospitality.

\end{document}